\documentclass[11pt]{article}
\usepackage[margin=1.1in]{geometry}

\usepackage{amsmath, amssymb, mathtools}
\usepackage{subcaption,graphicx}
\usepackage{xcolor}
\usepackage{booktabs}
\usepackage{array}

\usepackage[authoryear]{natbib}
\newtheorem{theorem}{Theorem}[section]
\newtheorem{lemma}{Lemma}

\usepackage{xurl}
\usepackage{kotex}
\usepackage{multirow}

\DeclareMathOperator*{\argmax}{argmax}

\newcommand{\V}{{\mathbb{V}ar}}
\newcommand{\C}{{\mathbb{C}ov}}

\newcommand{\lra}{\longrightarrow}

\newcommand{\bbE}{{ \mathbb{E}}}

\newcommand{\calC}{\mathcal{C}}

\newcommand{\calP}{\mathcal{P}}

\newcommand{\calU}{\mathcal{U}}

\newcommand{\kommentar}[1]{}
\def\Real{\hbox{I\kern-.1667em\hbox{R}}}
\def\Reals{\hbox{\scriptsize I\kern-.1667em\hbox{R}}}

\newcommand{\bsu}{\boldsymbol{u}}
\newcommand{\bsv}{\boldsymbol{v}}

\newcommand{\bsw}{\boldsymbol{w}}

\newcommand{\bss}{\boldsymbol{s}}

\newcommand{\bsV}{\boldsymbol{V}}

\newcommand{\bsB}{\boldsymbol{B}}
\newcommand{\bsS}{\boldsymbol{S}}

\newcommand{\diag}{\mbox{diag}}

\newcommand{\bfX}{{\bf X}}

\newcommand{\sg}{\Sigma}

\newcommand{\bea}{\begin{eqnarray*}}
	\newcommand{\eea}{\end{eqnarray*}}
\newcommand{\bean}{\begin{eqnarray}}
	\newcommand{\eean}{\end{eqnarray}}
\newcommand{\benu}{\begin{enumerate}}
	\newcommand{\eenu}{\end{enumerate}}

\newcommand{\bbR}{\mathbb{R}}
\newcommand{\bbN}{\mathbb{N}}
\newcommand{\bbP}{\mathbb{P}}

\newcommand{\bsg}{\boldsymbol{\sigma}}

\allowdisplaybreaks

\title{Scalable and optimal Bayesian inference for sparse covariance matrices via screened beta-mixture prior}
\author{
	Kyoungjae Lee\\Department of Statistics, Sungkyunkwan University
	\and
	Seongil Jo\footnote{Corresponding author.}  \\Department of Statistics,
	Inha University
	\and 
	Kyeongwon Lee\\ Department of Statistics, Seoul National University
	\and
	Jaeyong Lee \\
	Department of Statistics,
	Seoul National University
}

\begin{document}



\maketitle

\begin{abstract}
In this paper, we propose a scalable Bayesian method for sparse covariance matrix estimation by incorporating a continuous shrinkage prior with a screening procedure. In the first step of the procedure, the off-diagonal elements with small correlations are screened based on their sample correlations. In the second step, the posterior of the covariance with the screened elements fixed at $0$ is computed with the beta-mixture prior. The screened elements of the covariance significantly increase the efficiency of the posterior computation. The simulation studies and real data applications show that the proposed method can be used for the high-dimensional problem with the `large $p$, small $n$'. 
In some examples in this paper, the proposed method can be computed in a reasonable amount of time, while no other existing Bayesian methods can be. 
The proposed method has also sound theoretical properties. The screening procedure has the sure screening property and the selection consistency, and the posterior has the optimal minimax or nearly minimax convergence rate under the Frobeninus norm. 
\end{abstract}


\section{Introduction} \label{sec:intro}

Suppose we observe $n$ random samples $\bfX_n = (X_1,\ldots, X_n)^T \in \bbR^{n \times p}$ from the $p$-dimensional normal distribution:
\begin{equation}\label{model}
	X_i  \mid \sg \,\,\stackrel{iid}{\sim}\,\,  N_p\left( 0 , \Sigma\right), \quad i = 1, \ldots, n,
\end{equation}
where $n, p \in \bbN$, $\sg \in \calC_p$, $\calC_p$ is the set of all $p\times p$ positive definite matrices, and $\bbR$ and $\bbN$ are the sets of real numbers and integers, respectively. 
In this paper, we consider the high-dimensional setting where the number of variables $p = p_n$ can grow to infinity as $n\to\infty$.
For consistent estimation of $\Sigma$ in the high-dimensional setting, we reduce the number of parameters by assuming that most of the off-diagonal entries of the covariance matrix $\sg$ are zero, i.e., $\sg$ is $\ell_0$-sparse.

On the frequentist side, various penalties together with different types of loss or likelihood functions are applied to obtain sparsity in covariance estimation. 
\cite{bien2011sparse} optimized the nonconvex $l_0$-penalty together with normal likelihood with majorization and minimization algorithm and obtained a sparse covariance estimator. In estimating covariance matrix, often quadratic loss instead of the normal likelihood is used together with the $l_1$-penalty \citep{rothman2012positive, xue2012positive, maurya2016well}. 
\cite{lam2009sparsistency} derived the convergence rates under the Frobenius norm and sparsistency of the covariance and precision matrix estimators when the $l_1$-penalty is applied to the Cholesky factor, the precision matrix, or the covariance matrix. A procedure is said to be sparsistent if with probability converging to 1, the nonzero locations of the covariance elements are correctly estimated by the procedure.

Thresholding estimators of sparse covariances are proposed by \cite{bickel2008covariance}, \cite{el2008operator}, \cite{rothman2009generalized} and \cite{cai2011adaptive}. \cite{rothman2009generalized} proposed the generalized thresholding covariance estimator and obtained its sparsistency and convergence rate under the operator norm.
The generalized thresholding includes hard and soft thresholding, smoothly clipped absolute deviation (SCAD), and adaptive lasso. \cite{cai2011adaptive} derived the minimax convergence rate for sparse covariance matrices and showed that a thresholding estimator is minimax optimal adaptive to a wide range of sparse covariance classes. Since thresholding estimators do not always ensure positive definiteness, they need an 
adjustment for the positive definiteness.

Several Bayesian approaches for high-dimensional sparse covariance matrix estimation have also been developed, but with limited success.
\cite{silva2009hidden} and \cite{khare2011wishart} suggested the $G$-inverse Wishart prior and its variants, where $G$ indicates the covariance graph structure.
Although \cite{khare2011wishart} developed a block Gibbs sampler for posterior inference, it is only applicable when the covariance graph is decomposable.
Thus, the posterior inference is extremely challenging when the covariance graph is not decomposable, especially when $p$ is large.

For estimating general sparse covariance matrices, \cite{wang15} proposed the stochastic search structure learning (SSSL), which imposes spike and slab priors and exponential priors on off-diagonal and diagonal entries of a covariance matrix, respectively.
For posterior computation, the block Gibbs sampler cleverly circumventing the positive definiteness constraint was derived.
However, because the model space grows exceedingly huge when $p$ gets large, it has the practical issue that mixing of Markov chain is extremely slow.
Furthermore, the SSSL has to calculate the inverse and product of $(p-1)\times (p-1)$ matrices at each iteration of the block Gibbs sampler, which causes a high computational cost and frequent numerical errors.

In order to alleviate the slow mixing issue of the Markov chain, \cite{lee2021betamixture} proposed the beta-mixture shrinkage prior, which consists of continuous shrinkage priors and gamma priors for off-diagonal and diagonal entries of a covariance matrix, respectively.
A block Gibbs sampler was also constructed by appropriately modifying the algorithm in \cite{wang15}.
By replacing spike and slab priors with continuous shrinkage priors, the beta-mixture shrinkage prior drastically reduces the size of model space that the Markov chain should explore.
\cite{lee2021betamixture} empirically showed that the beta-mixture shrinkage prior actually achieves relative computational efficiency compared to the SSSL in terms of effective sample size.
However, the beta-mixture shrinkage prior also requires the calculation of the inverse and product of $(p-1)\times (p-1)$ matrices at each iteration of the block Gibbs sampler, which makes it problematic to employ even when $p$ is moderately large.
So far, no Bayesian approaches that are theoretically supported and practically viable for large covariance matrices have been developed.

In this paper, we propose a scalable Bayesian method for sparse covariance matrix estimation by incorporating a continuous shrinkage prior with a screening procedure based on sample correlation coefficients.
We call the resulting prior the screened beta-mixture shrinkage prior.
The key idea of the proposed prior is to reduce the dimension of the matrix to be handled in the block Gibbs sampler by discarding negligible components via an initial screening procedure.
As a consequence, the computational speed can be significantly enhanced while the frequency of numerical errors is reduced.


\begin{figure}[!tb]
	\centering
	\includegraphics[width=\textwidth]{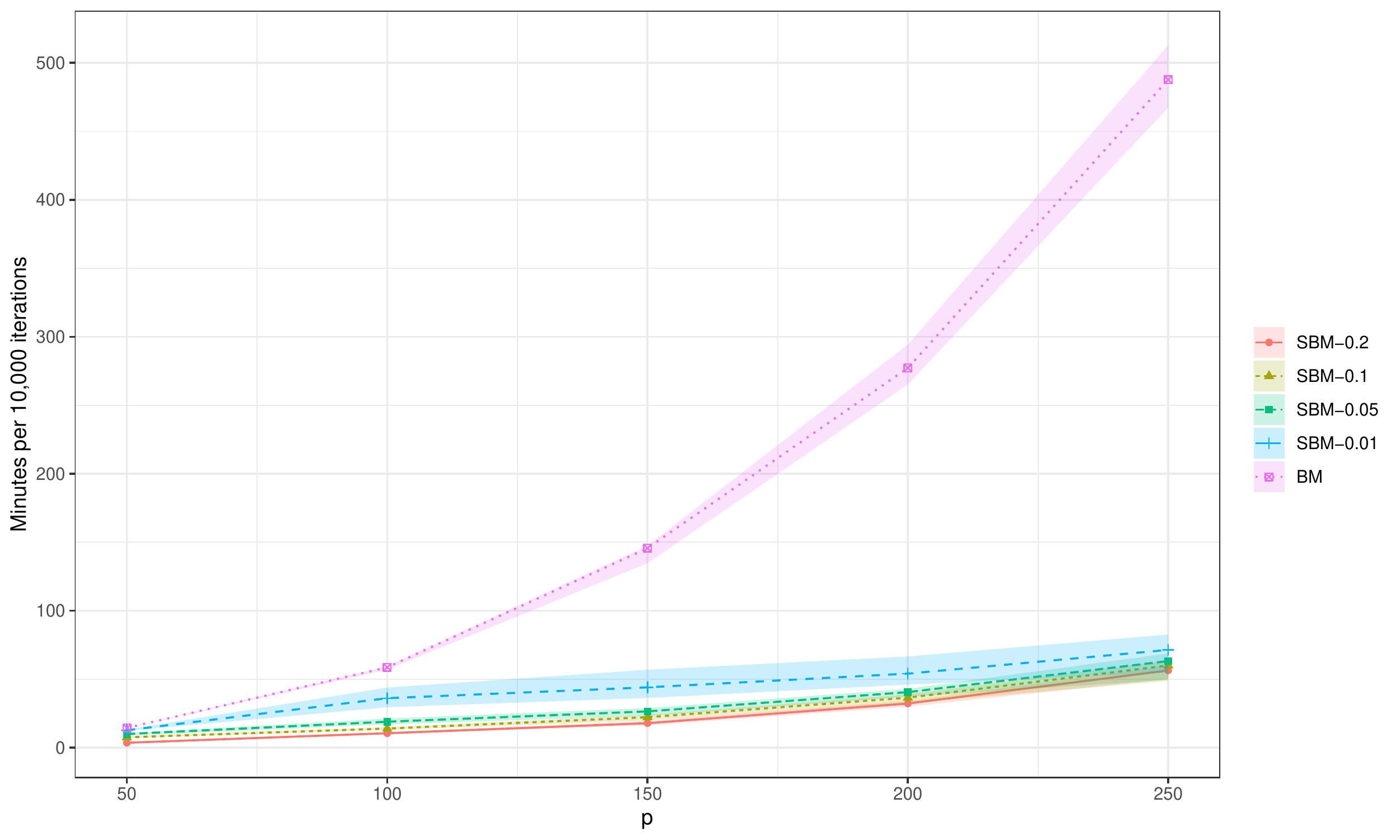}
	\caption{Computing times in minutes for 10,000 iterations of the beta-mixture shrinkage prior (BM) and the screened beta-mixture shrinkage prior (SBM) with $\alpha_{\rm FNR}$ = 0.01, 0.05, 0.1, and 0.2. The line indicates the median time, while the shaded area represents the 95\% interval between the 2.5\% and 97.5\% quantiles.}
	\label{fg:time_comp}
\end{figure}

Through simulation studies, we demonstrate that the proposed method can conduct substantially faster inference than the existing Bayesian methods.
Figure \ref{fg:time_comp} compares the computation times of the screened beta-mixture shrinkage prior and the beta-mixture shrinkage prior \citep{lee2021betamixture}, where the details are deferred to Sections \ref{subsec:speed} and \ref{subsec:choice_example}.
It shows that the screened beta-mixture shrinkage prior is much faster than the beta-mixture shrinkage prior, and especially for higher dimensions, the difference becomes more noticeable.
In Section \ref{sec:simulation}, we also empirically show that the performance of the proposed method is at least comparable to or better than the other existing Bayesian methods in terms of estimation while achieving substantial computational efficiency.

The screened beta mixture has sound theoretical justifications. 
We show that the proposed screening procedure has sure screening property (Theorem \ref{thm:sure_scr}) and selection consistency (Theorem \ref{thm:selection_cons}).
We also prove that the posterior convergence rate of the screened beta-mixture shrinkage prior under the Frobenius norm (Theorem \ref{thm:conv_rate}) is minimax or, at least, nearly minimax.

The rest of the paper is organized as follows.
In Section \ref{sec:screen}, we introduce a screening procedure via sample correlations.
In Section \ref{sec:cov_est}, we propose a screened beta-mixture shrinkage prior for sparse covariance matrix estimation and demonstrate the block Gibbs sampler for posterior inference. 
Furthermore, the computational efficiency of the proposed block Gibbs sampler is shown through simulation studies.
Theoretical properties of the proposed method are presented in Section \ref{sec:theory}, where the proofs are given in the Appendix.
A practical guide on the choice of hyperparameters is given in Section \ref{sec:hyperparameters}.
In Section \ref{sec:simulation}, we compare the performance of the proposed method with existing methods based on simulation studies, while a real data analysis is given in Section \ref{sec:realdata}.
We conclude with a discussion in Section \ref{sec:disc}.

\section{Screening via Sample Correlation Coefficients}\label{sec:screen}

\subsection{Screening procedure}\label{subsec:screening}
In this section, we introduce a screening procedure for sparse covariance matrices.
A screening procedure sets the elements of the covariance expected to be zero or very close to  zero. 
Several screening methods have been proposed for regression models, including sure independence screening (SIS), iterative SIS (ISIS) \citep{fan2008sure} and split-and-merge (SAM) approach \citep{song2015split}.
One of the main aims of screening methods is to decrease the amount of computation involved for statistical inference.

We propose a sample correlation-based screening method that reduces the number of nonzero off-diagonal entries by preserving only those with absolute sample correlation coefficients greater than a threshold. 
For a given threshold $r>0$ and $\sg =(\sigma_{jk}) \in \calC_p$, we define a screening procedure 
\[
	S_r(\sg) := \big\{ (j,k): 1\le j<k \le p, \,\, |\sigma_{jk}| > r \big\} .
\]
Suppose the data $\bfX_n = (X_1,\ldots, X_n)^T$ is centered, and let $\hat{R} = (\hat{\rho}_{jk})$ be the sample correlation matrix, where $\hat{\rho}_{jk} = \sum_{i=1}^n X_{ij}X_{ik} / \sqrt{ \sum_{i=1}^n X_{ij}^2 \sum_{i=1}^n X_{ik}^2 }$ for $1\le j , k\le p$.
A sample correlation-based screening for covariance matrices is then defined by
\begin{equation}\label{screening}
	S_r(\hat{R}) := \big\{ (j,k): 1\le j<k \le p, \,\, |\hat{\rho}_{jk}| > r \big\},
\end{equation}
which preserves only the off-diagonal indices with the corresponding absolute sample correlation is larger than a given threshold $r$.
Note that in regression models, the SIS method \citep{fan2008sure} also preserves the variables whose sample correlation with the response variable is relatively large, so the above screening procedure \eqref{screening} is reminiscent of the SIS.

We suggest using only the indices in \eqref{screening} for further statistical inference and considering the rest as zero.
By doing so, we effectively reduce the number of parameters, which leads to a substantial improvement in terms of computational speed.
The details of the inference on sparse covariance matrices after selecting off-diagonal entries by the screening method \eqref{screening} are given in Section \ref{sec:cov_est}.
We note here that the practical performance and computational speed of the proposed Bayesian method may depend on the threshold $r$.
In Section \ref{sec:hyperparameters}, we propose adaptive methods for the choice of $r$.

\subsection{Bayesian viewpoint of the proposed screening method}\label{subsec:another_view}

The proposed screening procedure \eqref{screening} can be viewed from the perspective of Bayesian hypothesis testing.
For any $1\le j < k \le p$, consider the hypothesis testing problem 
\begin{align}\label{H_jk}
	H_{0,jk}: \rho_{jk}=0  \text{ \, versus \, } H_{1,jk}: \rho_{jk}\neq 0 .
\end{align}
\cite{ly2016harold} derived Jeffreys' default Bayes factor \citep{jeffreys1998theory} for testing the nullity of $\rho_{jk}$.
Let $B_{10,jk,\kappa} (\bfX_n)$ be their Jeffreys' default Bayes factor using only the $j$th and $k$th columns of the data matrix $\bfX_n$.
They calculated a closed form expression of the Bayes factor as follows:
\begin{align*}
	B_{10,jk,\kappa} (\bfX_n) &= \frac{2^{\frac{\kappa-2}{\kappa}} \sqrt{\pi} }{\mathcal{B}\left(\frac{1}{\kappa}, \frac{1}{\kappa}\right) } \frac{ \Gamma\left(\frac{2+(n-1)\kappa}{2\kappa}\right) }{\Gamma\left(\frac{2+n\kappa}{2\kappa}\right) } {}_{2}F_1 \left(\frac{n-1}{2} , \frac{n-1}{2}; \frac{2+n\kappa}{2\kappa}; \hat{\rho}_{jk}^2 \right) ,
\end{align*}
where $\kappa>0$ is a hyperparameter, $\Gamma$ is a gamma function, $\mathcal{B}$ is a beta function, and ${}_{2}F_1$ is a hypergeometric function \citep{ly2016harold}.

Note that $B_{10,jk, \kappa} (\bfX_n)$ is an increasing function of $\hat{\rho}_{jk}^2$, and thus,
for any given $r>0$, there exists a unique constant $r_{\rm J} >0$ such that 
\begin{align}\label{screening_BF}
	S_r(\hat{R}) & \equiv 
	\left\{(j,k): \, 1\le j<k \le p, \, B_{10,jk, \kappa} (\bfX_n) > r_{\rm J} \right\} 
	=:  S_{\kappa, r_{\rm J} }(\hat{R})  .
\end{align}
Therefore, we can consider the screening procedure \eqref{screening} based on sample correlations as a method of selecting only the indices that reject the null hypothesis in \eqref{H_jk} based on Jeffreys' default Bayes factors.
This point of view is useful not only to justify our proposed screening method but also to select a threshold value for screening in Section \ref{subsec:r_choice}.

\section{Sparse Covariance Matrix Estimation}\label{sec:cov_est}

\subsection{Screened beta-mixture prior for covariance matrix}\label{subsec:trim_BM} 

We propose a shrinkage prior for the sparse covariance whose nonzero off-diagonal elements are obtained 
by the screening method \eqref{screening}.
We first define a prior distribution on sparse symmetric matrices with positive diagonal entries and normalize it on a space of positive definite matrices.

For some positive constants $\tau_1, a, b, c$ and $d$, let
\bean\label{GL_shrinkage}
\begin{split}
	\pi^u(\sigma_{jk} \mid \psi_{jk}) &=  N \left(\sigma_{jk} \,\Big|\,  0, \, \frac{\psi_{jk}}{1-\psi_{jk}} \tau_1^2 \right) ,   \\
	\pi^u(\psi_{jk}) & = {\rm Beta}(\psi_{jk} \mid a, \,\, b)  , \quad \text{ for } (j,k) \in S_r(\hat{R}), \\
	\pi^u(\sigma_{jj}) & = {\rm Gamma}(\sigma_{jj} \mid c, \,\, d), \quad \text{ for } j=1, \ldots, p,
\end{split}
\eean
where 
${\rm Beta}(a, b)$ is the beta distribution with shape parameters $a$ and $b$, and ${\rm Gamma}(c, d)$ is the gamma distribution with shape parameter $c$ and rate parameter $d$.  
The prior on each selected off-diagonal entry $\sigma_{jk}$ corresponds to the global-local shrinkage prior \citep{polson2010shrink}.
We then define a prior on symmetric matrices with the support $S_r(\hat{R})$ and positive diagonal elements as 
\bea
\pi_r^u(\Sigma) \,=\, \prod_{(j,k) \in \hat{S}_r } \Big\{  \int_0^1 \pi^u(\sigma_{jk} \mid  \psi_{jk})\pi^u(\psi_{jk}) d \psi_{jk} \Big\} 
\prod_{(j,k)\notin \hat{S}_r} \delta_0 (\sigma_{jk})
\prod_{j=1}^p \pi^u(\sigma_{jj}) I \big( \sg = \sg^T \big), 
\eea
where  $\hat{S}_r = S_r(\hat{R})$, $\delta_0$ is the Dirac measure at 0, and ``u'' stands for the unconstrained prior.

Now, we propose the prior for sparse covariance matrices by restricting $\pi_r^u(\sg)$ to the subspace of positive definite matrices: 
\bean\label{sprior}
{\pi}_r( \Sigma ) &=& \frac{\pi_r^u(\Sigma) I (\Sigma \in \mathcal{U}(\epsilon) ) }{ \pi_r^u( \Sigma \in \mathcal{U}(\epsilon)  ) } ,
\eean
where 
\bea
\mathcal{U}(\epsilon) &=& \Big\{  \sg\in \calC_p:  \epsilon < \lambda_{\min}(\sg)\le \lambda_{\max}(\sg) < \epsilon^{-1}      \Big\} 
\eea
for some constant $0<\epsilon<1$.  
We call  prior \eqref{sprior} the screened beta-mixture (SBM) shrinkage prior for covariance matrices.
Note that the beta-mixture shrinkage prior \citep{lee2021betamixture} can be seen as a special case of the SBM prior with $r=0$.

In the SBM prior \eqref{sprior}, we normalize the unconstrained prior $\pi_r^u(\Sigma)$ on $\mathcal{U}(\epsilon)$, a subset of $\calC_p$, instead of $\calC_p$ itself.
This is necessary for technical reasons to obtain desired asymptotic properties of posteriors: by restricting the support of the prior to $\mathcal{U}(\epsilon)$, we can efficiently control the eigenvalues of $\sg$ in the resulting posterior.
However, this restriction is not necessary in practice.
Throughout numerical studies, we found that the resulting posteriors from \eqref{sprior} with $\mathcal{U}(\epsilon)$ and $\mathcal{U}(0) \equiv \calC_p$ are indistinguishable as long as $\epsilon>0$ is sufficiently small. 
Therefore, we suggest using $\epsilon=0$ in practice to avoid computing eigenvalues of $\sg$ at each step.

We note here that the screening procedure does not need to rule out all zero off-diagonal entries because the SBM prior further shrinks the selected off-diagonal entries by using the global-local shrinkage prior \eqref{GL_shrinkage}.
Rather, it is desirable to identify a superset containing all of the true nonzero entries.
This reasoning will be applied later in Section \ref{subsec:r_choice} when choosing a threshold $r$.

\subsection{Posterior computation}\label{subsec:cov_post_comp}

We describe a posterior sampling algorithm for the proposed SBM prior. 
The algorithm is based on the block Gibbs sampler proposed by \cite{wang15} and \cite{lee2021betamixture}, but it can be much faster than the existing Bayesian methods because the screening procedure reduces the dimension of matrices used in the posterior computation.

Let $\bsV = (v_{jk}) \in \bbR^{p\times p}$ be a matrix of hyperparameters such that $v_{jk} = v_{kj} = \psi_{jk}\tau_1^2/(1-\psi_{jk})$ for $j < k$ and $v_{jk} = 0$ for $ j = k$.
To describe the algorithm in detail, we partition $\Sigma$, $\bsS = \bfX_n^T\bfX_n$ and $\bsV$ as follows:
\begin{equation*}
	\Sigma = \left(\begin{array}{cc}\Sigma_{11} & \bsg_{12} \\ \bsg_{12}^T & \sigma_{22} \end{array}\right), \quad 
	\bsS = \left(\begin{array}{cc}\bsS_{11} & \bss_{12} \\ \bss_{12}^T & s_{22} \end{array}\right), \quad
	\bsV = \left(\begin{array}{cc}\bsV_{11} & \bsv_{12} \\ \bsv_{12}^T & 0 \end{array}\right),
\end{equation*}
where $\Sigma_{11}, \bsS_{11} \in \calC_{p-1}$, $\bsV_{11}  \in \bbR^{(p-1)\times (p-1)}$, $\bsg_{12}, \bss_{12} ,\bsv_{12}  \in \bbR^{ p-1}$ and $\sigma_{22}, s_{22}  >0$.
Consider the change variables 
\begin{equation*}
	\left(\bsg_{12}, \sigma_{22}\right) \,\,\mapsto\,\, \left(\bsu = \bsg_{12}, \eta = \sigma_{22} - \bsg_{12}^T\Sigma_{11}^{-1}\bsg_{12} \right).	
\end{equation*}
Recall that, after applying the screening method, the resulting $\Sigma = ( \sigma_{jk} I(|\hat{\rho}_{jk}| > r ) )$ is a sparse covariance matrix.
Thus, $\bsg_{12}$ contains only a few nonzero entries that need to be updated.
Let $\bsu_{S_{\neq 0}} = \bsg_{12, S_{\neq 0}} = ( \sigma_{jp}  )_{j \in S_{\neq 0}}$, where $S_{\neq 0} = \{j : I(|\hat{\rho}_{jp}| > r)  =1 ,\,\, 1\le j \le p-1 \}$ is the set of indices of nonzero entries in $\bsg_{12}$.
Then, the conditional posterior distributions of $\bsu$, $\eta$ and corresponding hyperparameters can be calculated in closed forms as follows.
\begin{itemize}
	\item The conditional posterior of $\bsu$ is
	\bean\label{cond_post_u}
	&& \bsu_{S_{\neq 0}} \mid \eta, \bsv_{12, S_{\neq 0}}, \sg_{11}, \bfX_n  \nonumber\\
	&\sim& \ N_{|S_{\neq 0}| } \left[\left\{\bsB_{S_{\neq 0}} + \diag(\bsv_{12, S_{\neq 0}}^{-1})\right\}^{-1}\bsw_{S_{\neq 0}}, \,\,  \left\{\bsB_{S_{\neq 0}} + \diag(\bsv_{12,S_{\neq 0}}^{-1})\right\}^{-1}\right], \quad\,\,
	\eean
	where $\bsB_{S_{\neq 0}} = (\Sigma_{11}^{-1})_{(S_{\neq 0}, *)}\bsS_{11}  (\Sigma_{11}^{-1})_{(*, S_{\neq 0})} \eta^{-1} + \lambda  (\Sigma_{11}^{-1})_{(S_{\neq 0},S_{\neq 0})}$, $\bsv_{12, S_{\neq 0}} = (v_{jp})_{j \in S_{\neq 0} } $ and $\bsw_{S_{\neq 0}} =  (\Sigma_{11}^{-1})_{(S_{\neq 0} , *)} \bss_{12}\eta^{-1}$.
	Here, we denote $(\Sigma_{11}^{-1})_{(S_{\neq 0}, *)} \in \bbR^{|S_{\neq 0}| \times (p-1) }$ as a submatrix of $\Sigma_{11}^{-1}$ with the rows corresponding to $S_{\neq 0}$, where $(\Sigma_{11}^{-1})_{(S_{\neq 0},S_{\neq 0})} \in \bbR^{|S_{\neq 0}| \times |S_{\neq 0}|}$ and $(\Sigma_{11}^{-1})_{(*, S_{\neq 0})} \in \bbR^{(p-1)\times |S_{\neq 0}|}$ are defined similarly.

	\item The conditional posterior of $\eta$ is 
	\bea
	&& \eta \mid \bsu_{S_{\neq 0}}, \sg_{11}, \bfX_n  \\
	&\sim& {\rm GIG} \left(1 - n/2, \,\, \lambda, \,\, \bsu_{S_{\neq 0}}^T (\Sigma_{11}^{-1})_{(S_{\neq 0},*)}\bsS_{11} (\Sigma_{11}^{-1})_{(*, S_{\neq 0})} \bsu_{S_{\neq 0}} - 2 \bss_{12}^T (\Sigma_{11}^{-1})_{(*, S_{\neq 0})} \bsu_{S_{\neq 0}} + s_{22}\right) , 
	\eea
	where ${\rm GIG}(q, a,b)$ is the generalized inverse Gaussian distribution with the probability density function $f(x) \propto x^{q-1} e^{ -(ax +b/x) /2}, \,\, x>0$.
	
	\item For any $j \in S_{\neq 0}$, let $\phi_{jp} = \psi_{jp}/(1 - \psi_{jp})$. 
	By introducing an auxiliary variable $\zeta_{jp}$, the conditional posterior distributions of $\psi_{jp}$ and $\zeta_{jp}$ can be calculated as 
	\bea
	\phi_{jp} \mid \zeta_{jp}, \sigma_{jp} &\sim& {\rm GIG} \left(a - 1/2, \,\, 2\zeta_{jp},\,\, \sigma_{jp}^2/\tau_1^2\right)  , \\
	\zeta_{jp} \mid \phi_{jp} &\sim& {\rm Gamma} \left(a + b, \,\, \phi_{jp} + 1\right) .
	\eea
\end{itemize}

Details for deriving the above conditional posterior distributions can be found in \cite{lee2021betamixture}.
The only difference from \cite{lee2021betamixture} is that all the elements of $\bsu$ are exactly zero except for $\bsu_{S_{\neq 0}}$ part.
Note that by permuting columns and rows of $\Sigma$ and $\bsV$ in turn, one can obtain a Markov chain Monte Carlo (MCMC) sample of $\Sigma$ based on the above conditional posteriors.

\subsection{Computational efficiency of the SBM prior}\label{subsec:speed}

Note that $\sg_{11}^{-1} \in \bbR^{(p-1)\times (p-1)}$ can be easily updated at each iteration without directly calculating the inversion of $(p-1)\times(p-1)$ matrix. Let 
\begin{align*}
	\Sigma^{-1} = \left(\begin{array}{cc}
		\Omega_{11} & \boldsymbol{\omega}_{12} \\ 
		\boldsymbol{\omega}_{12}^T & \omega_{22} 
	\end{array}\right) ,
\end{align*}
then by the Woodbury matrix identity, we have $\sg_{11}^{-1} = \Omega_{11} - \boldsymbol{\omega}_{12} \boldsymbol{\omega}_{12}^T / \omega_{22}$.
Thus, after calculating $\sg^{-1}$ once, $\sg_{11}^{-1}$ can be calculated with computational complexity $O(p^2)$.

The crucial difference of the above block Gibbs sampler from that in \cite{lee2021betamixture} is the dimension of the conditional posterior of $\bsu$.
In \cite{lee2021betamixture}, $\bsu$ has $p-1$ nonzero entries with probability 1 in the conditional posterior.
The computational bottleneck of the Gibbs sampler in \cite{lee2021betamixture} is (a) multiplications of three $(p-1)\times (p-1)$ matrices $\Sigma_{11}^{-1} \bsS_{11} \Sigma_{11}^{-1}$ and (b) inversion of $(p-1)\times (p-1)$ matrix in the covariance matrix of $\bsu$.
Note that the total computational complexity for (a) and (b) is $O(p^3)$.
On the other hand, using the SBM prior, $\bsu$ contains only $|S_{\neq 0}|$ nonzero entries in the conditional posterior, and the main computational burden becomes the computation of $(\Sigma_{11}^{-1})_{(S_{\neq 0} *)}\bsS_{11}  (\Sigma_{11}^{-1})_{(*, S_{\neq 0})} $.
Its computational complexity is $O( p^2|S_{\neq 0}|)$ when $n \ge p$, but when $n<p$, it can be reduced to $O(np |S_{\neq 0}|)$ by calculating $(\Sigma_{11}^{-1})_{(S_{\neq 0}, *)} \bsS_{11} (\Sigma_{11}^{-1})_{(*, S_{\neq 0})} = n^{-1}( \bfX_{n,1} (\Sigma_{11}^{-1})_{(*, S_{\neq 0})})^T \bfX_{n,1} (\Sigma_{11}^{-1})_{(*, S_{\neq 0})}$,where $\bfX_{n,1}\in \bbR^{n\times (p-1)}$ is the submatrix of $\bfX_n$ without the $p$th column.
Thus, the SBM prior enables fast posterior computation especially when $\hat{S}_r$ is sparse so that $|S_{\neq 0}| \ll p$.
We further describe the empirical computational efficiency of the proposed SBM prior through a brief simulation study in Section \ref{subsec:choice_example}.

\section{Theoretical Results}\label{sec:theory}

\subsection{Screening properties}\label{subsec:screen_properties}

When a screening procedure is applied, one of the biggest concerns is whether it has the sure screening property, i.e., whether  it keeps all the significant variables with probability tending to $1$ \citep{fan2008sure}.
If a screening method has the sure screening property, we call it  a sure screening method.
In this section, we show that the proposed sample correlation-based screening procedure in Section \ref{sec:screen} achieves the sure screening property and selection consistency under mild conditions.

Let $\Sigma_0 = (\sigma_{0,jk})$ and $S_0 = S_0(\Sigma_0) := \{ (j,k) : 1\le j<k\le p, \,\, |\sigma_{0,jk}| \neq 0 \}$ be the true covariance matrix and its support, respectively.
Let $R_0 = (\rho_{0,jk})$ be the true correlation matrix.
The following condition provides a lower bound for the size of nonzero entries of the true correlation matrix $R_0$.
\begin{enumerate}
	\item[\bf (A1)] There exist a constant $\delta>0$ such that, for any $(j,k) \in S_0$ and some $r>0$,
	\bea
	|\rho_{0,jk} |    - r    &\ge& ( 2 + \delta) \sqrt{ \frac{(1+\rho_{0,jk}^2) \log(n\vee p)}{n} } .
	\eea
\end{enumerate}

We call condition (A1) the beta-min condition.
When $r=0$, this condition roughly implies that the minimum absolute nonzero entry in $R_0$ is 
larger than $2 \sqrt{ \log (n\vee p) /n}$.
Similar conditions have been used to guarantee consistent selections for nonzero entries in regression coefficient vectors \citep{castillo2015bayesian,martin2017empirical,lee2020bayesian} 
and covariance matrices \citep{lam2009sparsistency,cai2011adaptive}.
Under condition (A1), the following theorem shows that $S_r(\hat{R})$ has the sure screening property as long as $0 < r \le 1/2$.

\begin{theorem}[Sure screening]\label{thm:sure_scr}
	Under model \eqref{model}, assume condition (A1) holds.
	If $0<r \le 1/2$, $n>2$ and $|S_0(\sg_0)| \le s_0$,
	\bea
	\bbP_0 \big\{  S_0(\sg_0) \subseteq S_r( \hat{R}) \big\} &\ge& 1 - \frac{23s_0 (nr \wedge 1) }{(2+\delta)\sqrt{\log(n\vee p)} } (n\vee p)^{-2(1+\delta)} ,
	\eea
	which implies $\bbP_0 \big\{  S_0(\sg_0) \subseteq S_r( \hat{R}) \big\} \to 1$ as $n\to\infty$.
\end{theorem}

Although the sure screening property in Theorem \ref{thm:sure_scr} is enough to guarantee that all the nonzero entries would be selected, one might be interested in an exact support recovery of the true covariance matrix.
For that, the threshold $r$ should be carefully chosen so that it can successfully distinguish the noises (zero entries) from the signals (nonzero entries).

\begin{enumerate}
	
	\item[\bf (A2)] $r = C_{\rm th} \sqrt{\log (n\vee p)  /n }$ for some constant $C_{\rm th}>0$.
\end{enumerate}

Condition (A2) implies that threshold $r$ should be at least of order $\sqrt{\log(n \vee p) /n}$ to detect all the true zero entries in $\Sigma_0$.
Note that it also affects the lower bound for nonzero entries in $\Sigma_0$ due to condition (A1).
In fact, condition (A1) together with (A2) can be further simplified.
If condition (A2) holds,  a sufficient condition for (A1) is that
there exists a constant $\delta>0$ such that 
\bean \label{eq:A1-1}
|\rho_{0,jk} |   &\ge& ( 2 + C_{\rm th}  + \delta) \sqrt{ \frac{ 2 \log(n\vee p)}{n} }  \quad \text{ for any } (j,k) \in S_0 .  \label{simple_betamin}
\eean
However, this is an unnecessarily strong condition, so we use conditions (A1) and (A2) which are slightly more complicated than \eqref{eq:A1-1}. 

Theorem \ref{thm:selection_cons} shows that the sure screening $S_r(\hat{R})$ can exactly recover the true support of $\Sigma_0$ with probability tending to $1$ under mild conditions, i.e., it has selection consistency.

\begin{theorem}[Selection consistency]\label{thm:selection_cons}
	Under the conditions in Theorem \ref{thm:sure_scr} and (A2),
	\bea
	\bbP_0 \big\{  S_0(\sg_0) = S_r (\hat{R}) \big\} &\ge& 1  - \frac{23s_0 (nr \wedge 1) }{(2+\delta)\sqrt{\log(n\vee p)} } (n\vee p)^{-2(1+\delta)}  \\
	&&-\,\, \frac{2\{p(p-1) - 2s_0 \}  }{C_{\rm th}\sqrt{\log (n\vee p)} } (n\vee p)^{- \frac{C_{\rm th}^2}{2} } .
	\eea
	Furthermore, if $C_{\rm th} \ge 2$, it implies $\bbP_0 \big\{ S_0(\sg_0) = S_r (\hat{R})\big\}  = 1 -o(1)$ as $n\to\infty$.
\end{theorem}

It is worth comparing the required conditions in this paper with those in \cite{cai2011adaptive}.
\cite{cai2011adaptive} used a beta-min condition 
\bean\label{cai_betamin}
|\sigma_{0,jk}| &\ge& (2 + C_{\rm th}  + \delta ) \sqrt{ \frac{ \theta_{jk} \log(n\vee p)}{n} }  \quad \text{ for any } (j,k) \in S_0
\eean
for some $\delta>0$, where $\theta_{jk} = \sigma_{0,jj} \sigma_{0,kk} + \sigma_{0,jk}^2$ under the Gaussian model.
By dividing  both sides of \eqref{cai_betamin} by $\sqrt{\sigma_{0,jj} \sigma_{0,kk}}$,  we obtain
\bean\label{simple_betamin2}
|\rho_{0,jk}| &\ge& (2 + C_{\rm th}  + \delta ) \sqrt{ \frac{ (1+ \rho_{0,jk}^2)\log(n\vee p)}{n} }  \quad \text{ for any } (j,k) \in S_0 .
\eean
Note that with condition (A2), \eqref{simple_betamin2} is a sufficient condition for (A1).
Thus, the beta-min condition used in this paper is slightly weaker than that used in \cite{cai2011adaptive}.

\subsection{Posterior convergence rate}\label{subsec:cov_theory}

In this section, we show the posterior convergence rate of the SBM prior.
It turns out that the SBM prior achieves the minimax or nearly minimax rate for sparse covariance matrices under the Frobenius norm.
Therefore, the proposed method is not only computationally scalable but also optimal in terms of posterior convergence rate.

For a given integer $0\le s_0\le p(p-1)/2$ and a real number $0<\epsilon_0 <1$, we define the parameter space for sparse covariance matrices,
\bea 
\mathcal{U}(s_0, \epsilon_0)
&=& \Big\{ \Sigma \in \calC_p: |S_0(\sg)| \le s_0, \,\,  \sg \in \calU(\epsilon_0) \Big\} . 
\eea 
To derive posterior convergence rate, we introduce the following conditions.

\begin{enumerate}
	\item[\bf (A3)] $\sg_0 \in \calU(s_0,\epsilon_0)$ for constant $0< \epsilon_0 <1$ and some integer $0\le s_0\le p(p-1)/2$ such that $(p+s_0)\log p =o(n)$ and $p \asymp n^{\beta}$ for some $0<\beta<1$.  
	\item[\bf (A4)]  The hyperparameters satisfy  
	$a=b=1/2, c=1, d= \lambda /2$, 
	$0< \epsilon   \le \max(1/3, \epsilon_0)$
	and $(p^2 \sqrt{n})^{-1}\lesssim  \tau_1 \lesssim \sqrt{ s_0 \log p } \,(p^2 \sqrt{n})^{-1}$, where $\lambda = O(1)$ and $\epsilon = O(1)$.
\end{enumerate}

Condition (A3) means that the true covariance matrix $\Sigma_0$ is a sparse matrix with at most $s_0$ nonzero entries in its lower triangular part, where $s_0 \log p = o(n)$.
It also assumes that $p$ has the same rate with $n^\beta$ for some $0<\beta<1$, which implies that the number of variables $p$ should be smaller than the sample size $n$.
Similar conditions have been used in the literature to derive convergence rates for covariance matrices under the Frobenius norm.
For examples, see \cite{lam2009sparsistency}, \cite{liu2019empirical} and \cite{lee2021betamixture}.

Condition (A4) is the conditions for hyperparameters in the SBM prior. 
The choice $a=b=1/2, c=1$ and $d= \lambda /2$ implies that we use the horseshoe prior with the global shrinkage parameter $\tau_1$ and the exponential prior with the rate parameter $\lambda$ for $\pi^u(\sigma_{jk})$ and $\pi^u(\sigma_{jj})$, respectively.
By assuming $(p^2 \sqrt{n})^{-1}\lesssim  \tau_1 \lesssim \sqrt{ s_0 \log p } \,(p^2 \sqrt{n})^{-1}$, we require the global shrinkage parameter $\tau_1$ should be sufficiently small.
\cite{song2017nearly} used a similar condition to obtain theoretical results.
A practical suggestion for the choice of $\tau_1$ will be described in Section \ref{sec:hyperparameters}.

For any $A = (a_{jk}) \in \bbR^{p \times p}$, the Frobenius norm is defined as $\|A \|_F = (\sum_{j=1}^p \sum_{k=1}^p a_{jk}^2 )^{1/2}$.
The next theorem shows the posterior convergence rate of the induced posterior under the Frobenius norm.

\begin{theorem}\label{thm:conv_rate}
	Under model \eqref{model} and prior \eqref{sprior}, assume conditions (A1), (A3), (A4) and $0< r\le 1/2$ hold.
	Then, 
	\bea
	\pi_r \Big\{  \|\sg - \sg_0 \|_F^2 \ge M \frac{(p+s_0)\log p}{n}  \mid \bfX_n  \Big\}  &\lra& 0 ~ \text{ in $\bbP_{0}$-probability}
	\eea
	for some large constant $M>0$, as  $n\to\infty$.
\end{theorem}

By Theorem 3.2 in \cite{lee2021betamixture}, the obtained posterior convergence rate in Theorem \ref{thm:conv_rate} is minimax or, at least, nearly minimax depending on the relationship between $s_0$ and $p$.
Specifically, it coincides with the minimax rate if $3p < s_0 < p^{3/2 - \epsilon}$ for some small constant $\epsilon>0$.
Even when $3p < s_0 < p^{3/2 - \epsilon}$  does not hold, the obtain posterior convergence rate is equal to the minimax lower bound up to $\log p$ factor.
Therefore, Theorem \ref{thm:conv_rate} says that the SBM prior achieves the optimal or, at least, nearly optimal posterior convergence rate in terms of minimaxity.

\section{Choice of hyperparameters}\label{sec:hyperparameters}

In this section, we provide a practical choice of hyparameters in the SBM prior.
To satisfy condition (A4), we first set $a=b=1/2$, $c=1$ and $d= \lambda/2$ with $\lambda=1$.
As mentioned before, a constant $0<\epsilon<1$ is introduced in \eqref{sprior} due to technical reasons.
However, as long as it is sufficiently small, the value of $\epsilon$ does not affect practical performance of the SBM prior.
In the numerical study, we use $\epsilon = 0$ for computational convenience.
Note that this choice leads to $\calU(\epsilon) = \calC_p$ in \eqref{sprior}.

\subsection{Choice of global shrinkage parameter $\tau_1$}\label{subsec:tau1}

The choice of the global shrinkage parameter $\tau_1$ affects the number of nonignorable off-diagonal elements in resulting covariance matrices.
To satisfy condition (A4), $(p^2 \sqrt{n})^{-1} \lesssim \tau_1 \lesssim \sqrt{s_0 \log p}\, (p^2 \sqrt{n})^{-1}$ should hold.
However, we found that $\tau_1 = (p^2 \sqrt{n})^{-1}$ is too small so that this overly shrinks nonzero entries in our simulation study.
On the other hand, the choice $\tau_1 = \sqrt{s_0\log p} \, (p^2 \sqrt{n})^{-1}$ is not available in practice because it depends on the unknown sparsity $s_0$.
Therefore, to circumvent these issues, we assume that approximately only a fixed (small) percentage of off-diagonal entries are nonzero, i.e., $s_0 \asymp p^2$, and suggest using $\tau_1 = \sqrt{\log p} \, (p \sqrt{n})^{-1}$, which satisfies condition (A4) under $s_0 \asymp p^2$.
We use $\tau_1 = \sqrt{\log p} \, (p \sqrt{n})^{-1}$ in the subsequent numerical study.

\subsection{Choice of threshold $r$}\label{subsec:r_choice}

A threshold $r$ for the screening $S_r(\hat{R})$ determines the number of nonzero off-diagonal entries in the SBM prior.
Although we provide theoretical conditions for guaranteeing asymptotic properties of the screening procedure, they might be too restrictive given $n$ and $p$ in practice. 
For example, when $(n,p)=(100, 300)$, we have $2 \sqrt{\log (n\vee p) /n} \approx 0.4777$.
Note that to guarantee selection consistency (Theorem \ref{thm:selection_cons}), we need to choose $r= C_{\rm th} \sqrt{\log (n\vee p) /n}$ with $C_{\rm th} \ge 2$, which leads to $r \ge 0.4777$.
Suppose that we choose the threshold $r= 0.4777$.
Then, the resulting screening procedure will select only off-diagonal elements with sample correlations larger than $0.4777$ in magnitude, which may miss a lot of true nonzero elements.

In this section, we suggest two adaptive methods for selecting a threshold $r$. 
As a simple choice, a quantile of absolute sample correlation coefficients can be used as a threshold $r$. 
For example, when the true covariance matrix is expected to be very sparse, the $0.2$-quantile of absolute sample correlation coefficients would be a reasonable choice.
Otherwise, if there is a prior knowledge on the proportion of nonzero entries, it could be used to choose an appropriate quantile.
The advantages of this method are that it is easy and fast to perform, and the calculation speed can be improved as much as desired.

Alternatively, there is a more sophisticated way to choose the threshold $r$ using Jeffreys' Bayes factors.
Using the relationship \eqref{screening_BF}, we select a cutoff $r_{\rm J}$ for Jeffreys' Bayes factors instead of a threshold $r$ for sample correlations. 
We suggest using a false negative rate (FNR)-based approach, which empirically controls the FNR at a prespecified level based on simulated datasets.
We generate $B$ simulated datasets independently, say $\{ \bfX_n^{(1)}, \ldots, \bfX_n^{(B)} \}$, where one dataset consists of $n$ random samples from a bivariate normal with mean zero and correlation matrix with a nonzero correlation $\rho_{\star} >0$.
Here, $\rho_{\star}$ corresponds to a lower bound of {\it meaningfully large} correlations, which means that correlations less than $\rho_{\star}$ in magnitude will be considered as indistinguishable from zero.
It could be chosen based on a prior knowledge or the characteristics of the data at hand.
We calculate Jeffreys' default Bayes factors for each dataset and conclude that the correlation is nonzero if the Bayes factor is larger than a cutoff $r_{\rm J}$.
Since the data-generating correlation $\rho_{\star}$ is nonzero, the Bayes factors smaller than $r_{\rm J}$ correspond to false negatives.
Therefore, we can control the FNR at the prespecified level $\alpha_{\rm FNR}$ by choosing a cutoff $r_{\rm J}$ as the $\alpha_{\rm FNR}$-quantile of Jeffreys' default Bayes factors.

\begin{figure}[!tb]
	\centering
	\includegraphics[width=\textwidth]{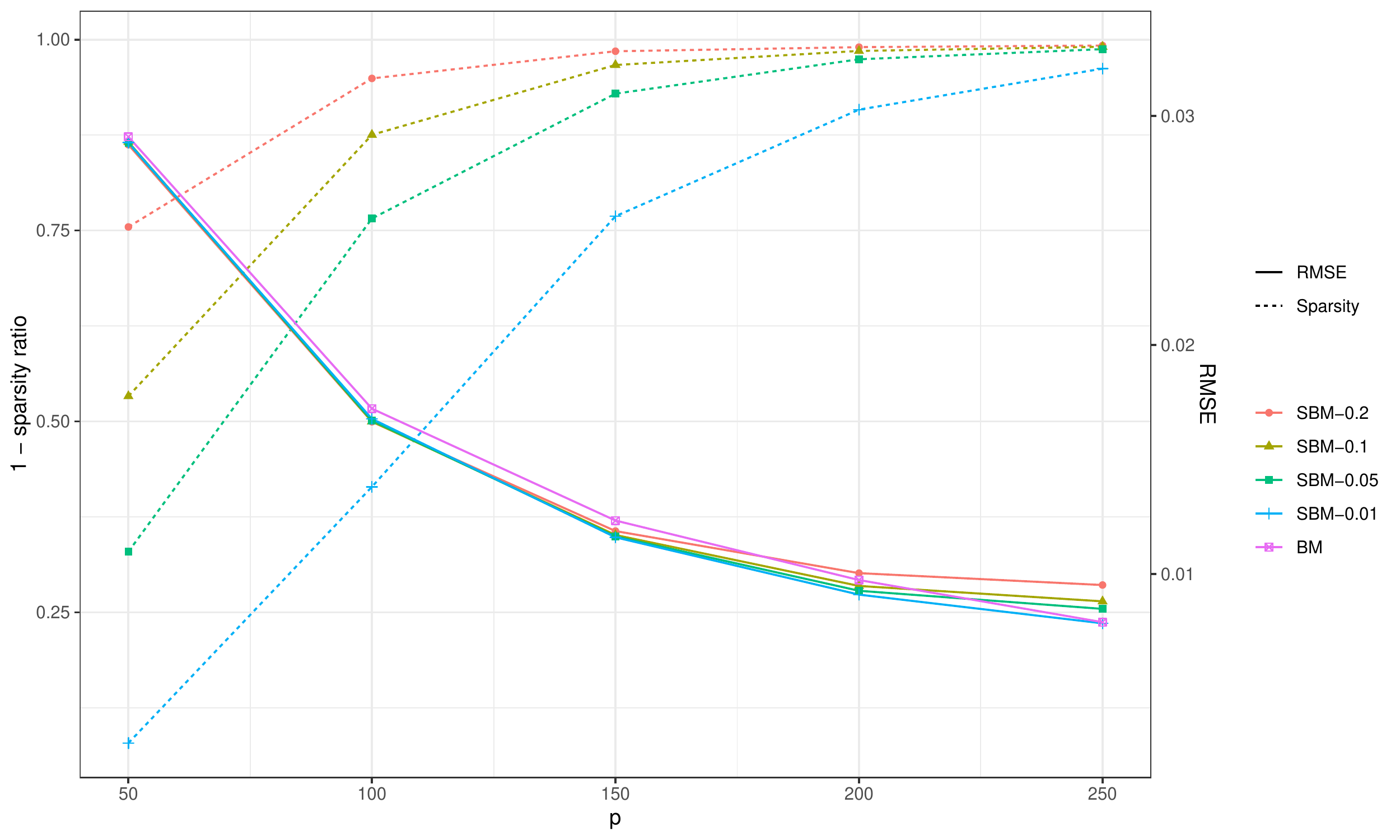}
	\caption{$1-$sparsity ratio (dashed lines) and \texttt{rmse} (solid lines) for beta-mixture shrinkage prior (BM) and screened beta-mixture shrinkage prior (SBM) with $\alpha_{\rm FNR} =0.01,~0.05,~0.1$ and $0.2$.}
\label{fg:sparsity_comp}
\end{figure}

\begin{figure}[!ht]
\centering
    \begin{subfigure}{0.95\textwidth}
    \centering
    \includegraphics[width=\textwidth]{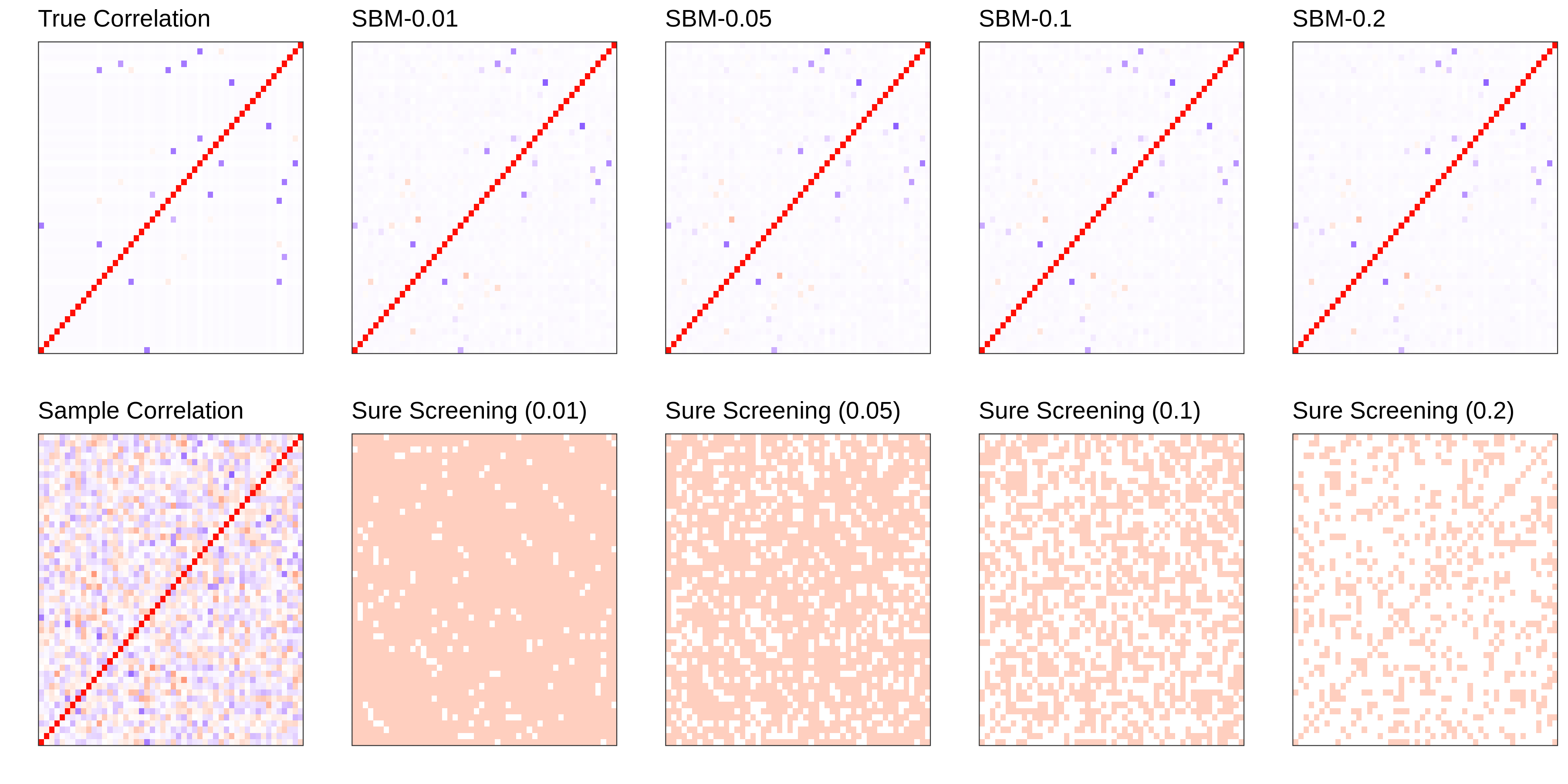}
    \caption{$p=50$}
    \end{subfigure}
    \begin{subfigure}{0.95\textwidth}
    \centering
    \includegraphics[width=\textwidth]{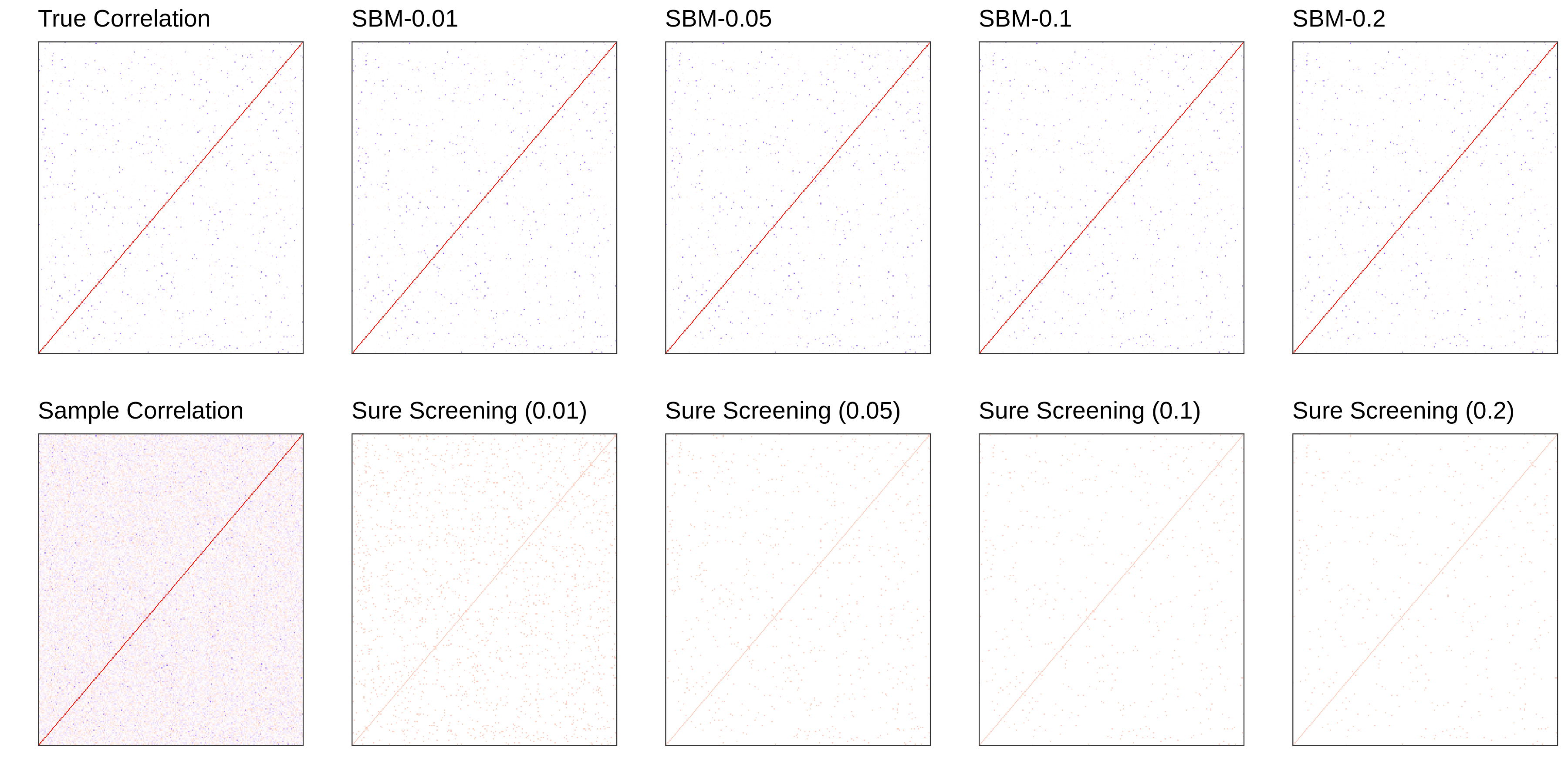}
    \caption{$p=250$}
    \end{subfigure}
    \caption{Correlation matrices and sure screening results for $p=50$ and $250$. The second to fifth columns correspond to the results when $\alpha_{\rm FNR}$ is $0.01$, $0.05$, $0.1$ and $0.2$, respectively.}
\label{fg:vis_cor}
\end{figure}

\subsection{Examples}\label{subsec:choice_example}

We compare computational efficiency of the beta-mixture prior and SBM prior when $p\in \{50, 100, 150, 200, 250\}$.
For each $p$, $n=2p$ observations $X_1,\ldots, X_n$ were generated from $N_p(0, \sg_0)$, where $\sg_0 = (\sigma_{0,jk})$ is the true covariance matrix.
We chose 1\% among lower triangular entries of $\sg_0$ and sampled their values from ${\rm Unif}(0.1, 0.4)$ independently, while the remaining lower triangular entries are fixed to zero.
We set $\sigma_{0,jj}= 1$ for all $j$.
When $\sg_0$ is not positive definite, we make it positive definite by adding $\{-\lambda_{\min}(\sg_0) + 0.1^5\} I_p$ to $\sg_0$ for numerical stability.

For the choice of threshold $r$, various $\alpha_{\rm FNR} \in \{0.01,~ 0.05,~ 0.1,~ 0.2\}$ are considered. For each method, other hyperparameters were set as described in Section \ref{sec:simulation}. 
The computation times for obtaining 10,000 posterior samples are summarized in Figure \ref{fg:time_comp} in Section \ref{sec:intro}.
Note that when $p=250$, the beta-mixture prior took nearly 8 hours to obtain 10,000 samples, which prevents using it in high-dimensional settings. In contrast, the SBM prior took approximately 1 hour.
Therefore, the SBM enables scalable inference even in high-dimensional situations that the beta-mixture prior could not handle.

We compared the sparsity ratio, which represents the proportion of nonzero off-diagonal entries, to demonstrate the computational efficiency of the methods. Figure \ref{fg:sparsity_comp} illustrates the comparison between the sparsity ratio and the root mean squared error (\texttt{rmse}) of the SBM prior and the beta-mixture shrinkage prior. 
The lines represent the median value of 50 repeated experiments. Note that the sparsity ratio decreases as $\alpha_{\rm FNR}$ increases. 
The sparsity ratio also decreases as $p$ grows, probably because we set $n=2p$ and the screening performance improves as $p$ increases.
In Figure \ref{fg:vis_cor}, the correlation matrix patterns and sure screening results are presented. With the increase in $\alpha_{\rm FNR}$, the SBM estimate results in a sparser covariance. Note that regardless of the $\alpha_{\rm FNR}$ value, the estimated covariance matrix exhibits a similar sparsity pattern. In other words, even the unscreened components are estimated to be close to zero during the estimation, providing high accuracy. Choosing a small $\alpha_{\rm FNR}$ has the advantage of reducing false negative errors but may impact computation time.
As shown in Figure \ref{fg:sparsity_comp}, the smallest value of $\alpha_{\rm FNR}$, which is 0.01, demonstrated accurate results. However, for sufficiently large values of $n=2p$, small $\alpha_{\rm FNR}$ yielded efficient results in Figure \ref{fg:time_comp}.

In the numerical study in Section \ref{sec:simulation}, we use the FNR-based approach to choose a threshold.
The hyperparameter $\kappa=1$ is chosen as suggested by \cite{ly2016harold}.
The size of meaningful correlations and the target FNR set at $\rho_{\star}=0.2$ and $\alpha_{\rm FNR}=0.01$, respectively.
We use \verb|correlationBF| function in the \textsf{R} package \verb|BayesFactor| to calculate the Jeffreys' default  Bayes factor.

\section{Numerical Study}\label{sec:simulation}
In this section, we compare the performance of the SBM prior \eqref{sprior} to other existing sparse covariance estimation methods: stochastic search structure learning (SSSL) \citep{wang15}, beta-mixture (BM) shrinkage prior  \citep{lee2021betamixture}, Stein-type shrinkage (STS) \citep{touloumis15}, adaptive thresholding (AT) \citep{cai2011adaptive} and sample covariance (SC).
Among them, SSSL and BM are Bayesian approaches, while STS and AT are frequentist approaches. 

For the AT method, we fix a shrinkage parameter for the adaptive lasso rule with value 4 and choose a tuning parameter for thresholding by cross-validation with values on $[0, 4]$. For SSSL, we set variances of the prior for off-diagonal elements with $v_0 = 0.02$ and $v_1 = 1$, the hyperparameter of the diagonal elements with $\lambda = 1$, and select the prior inclusion probability with $\pi = 2/(p-1)$.
We generate 4,000 posterior samples by using the block Gibbs sampler given in Section \ref{sec:cov_est} and compute posterior means with the last 2,000 samples among posterior samples to estimate the covariance matrix for Bayesian approaches. We use the \textsf{R} package {\tt ShrinkCovMat} \citep{touloumis15} to implement the STS method and the \textsf{R} code obtained from \url{https://github.com/qpsy/hotelhd} for the AT method. 
We provide an \textsf{R} package {\tt bspcov} (available at \url{https://github.com/statjs/bspcov}) for implementation of the proposed SBM prior and BM prior. 
All methods are evaluated with Linux clusters with Intel Xeon processor.

We generate $n$ random samples from the $p$-dimensional multivariate normal distribution with the zero mean vector and the covariance matrix $\Sigma_0 = (\sigma_{0, jk})$. 
For the covariance matrix $\Sigma_0$, we set the diagonal elements to be one and consider the below three scenarios to generate the off-diagonal elements.
Note that we add a small value $0.1^5 - \lambda_{\min}(\Sigma_0)$ to diagonal elements of a generated covariance matrix $\Sigma_0$ unless it is positive definite.
\begin{itemize}
	\item {\bf Random}. 
	Each off-diagonal element follows ${\rm Unif}(-0.8, -0.2)$ with probability $1/p$ and is set to be zero with probability $1-1/p$.
	
	\item {\bf Hubs}. The rows/columns are partitioned into disjoint groups $\{G_k\}_{k=1}^K$. Each group $G_k$ has a hub member $h_k \in G_k$ such that $\sigma_{0, i h_k}\sim {\rm Unif}(0.1, 0.5)$ for $i \in G_k$ and $\sigma_{0, ij} = 0$ otherwise. 
	We set $K = 10$.
	
	\item {\bf Cliques}. The rows/columns are partitioned into disjoint groups $\{G_k\}_{k=1}^K$.
	In each group, we choose $m$ members, say $G_k^m$, and set
	$\sigma_{0, ij} \sim {\rm Unif}(-0.45, -0.1)$ for $i, j \in G_k^m$ and $i\neq j$,
	while the others are set to be zero. 
	Here we choose $K = 10$ and $m = 3$.
\end{itemize}

To evaluate the estimation performance of an estimator $\widehat{\sg} = (\hat{\sigma}_{jk} )$, we compute two accuracy measures: i) the root mean squared error ({\tt rmse}), $\|\widehat{\Sigma} - \Sigma_0\|_F / p$,
and ii) the element-wise maximum norm ({\tt mnorm}), $\|\widehat{\sg} - \sg_0\|_{\max}= \mbox{max}_{1\le j \le k\le p} |\hat{\sigma}_{jk} - \sigma_{0,jk}|$. 
The averages and standard deviations of each accuracy measure are calculated over 50 replications. 

\subsection{Fixed $n$ and varying $p$}\label{sec:simulation1}
In this subsection, we evaluate the performance of the proposed SBM prior in comparison to other competitors. The comparison is conducted using a fixed sample size of $n=150$, while the dimension varies within the range of $p \in \{100, 200, 300\}$.

\begin{table}[!tb]
\caption{\label{tab:result1}Averages and standard deviations (in parentheses) of \texttt{rmse} and \texttt{mnorm} when $n = 150$ and $p = 100$.}
\centering
\fontsize{9}{11}\selectfont
\begin{tabular}[t]{llcccccc}
\toprule
Structures & Measures & SBM & BM & SSSL & STS & AT & SC\\
\midrule
 &  & \textbf{0.0177} & 0.018 & 0.0196 & 0.044 & 0.0244 & 0.0951\\

 & \multirow[t]{-2}{*}{\raggedright\arraybackslash \texttt{rmse}} & (0.0013) & (0.0014) & (0.0015) & (0.0004) & (0.0017) & (0.002)\\

 &  & 0.4124 & 0.417 & 0.4405 & 0.6294 & 0.4679 & \textbf{0.4096}\\

\multirow[t]{-4}{*}{\raggedright\arraybackslash \textbf{Random}} & \multirow[t]{-2}{*}{\raggedright\arraybackslash \texttt{mnorm}} & (0.0549) & (0.0533) & (0.068) & (0.0243) & (0.042) & (0.0486)\\
\cmidrule{1-8}
 &  & \textbf{0.0159} & \textbf{0.0159} & 0.0172 & 0.0278 & 0.0181 & 0.0824\\

 & \multirow[t]{-2}{*}{\raggedright\arraybackslash \texttt{rmse}} & (0.0011) & (0.0012) & (0.0011) & (0.0005) & (0.0014) & (0.0015)\\

 &  & \textbf{0.3348} & 0.341 & 0.3481 & 0.4494 & 0.3988 & 0.3409\\

\multirow[t]{-4}{*}{\raggedright\arraybackslash \textbf{Hubs}} & \multirow[t]{-2}{*}{\raggedright\arraybackslash \texttt{mnorm}} & (0.0476) & (0.0459) & (0.0473) & (0.0132) & (0.0414) & (0.0353)\\
\cmidrule{1-8}
 &  & \textbf{0.017} & 0.0174 & 0.0188 & 0.0235 & 0.0193 & 0.0824\\

 & \multirow[t]{-2}{*}{\raggedright\arraybackslash \texttt{rmse}} & (0.0012) & (0.0012) & (0.0013) & (0.0005) & (0.001) & (0.0015)\\

 &  & \textbf{0.3447} & 0.353 & 0.3717 & 0.416 & 0.3838 & 0.3475\\

\multirow[t]{-4}{*}{\raggedright\arraybackslash \textbf{Cliques}} & \multirow[t]{-2}{*}{\raggedright\arraybackslash \texttt{mnorm}} & (0.0696) & (0.0672) & (0.0753) & (0.0269) & (0.0417) & (0.0491)\\
\bottomrule
\end{tabular}
\end{table}

Table \ref{tab:result1} shows the results for the moderate high-dimensional setting, $p = 100$. 
The value in parentheses represents the standard deviation for each measure. 
Overall, the SBM prior works reasonably well in terms of {\tt rmse} compared to other competitors for all covariance structures we consider.
Among Bayesian methods, continuous shrinkage prior-based methods, SBM and BM priors, tend to outperform the spike and slab prior-based method, the SSSL approach.

\begin{table}[!tb]
\caption{\label{tab:result2}Averages and standard deviations (in parentheses) of \texttt{rmse} and \texttt{mnorm} when $n = 150$ and $p \in \{200,300 \}$}
\centering
\fontsize{9}{11}\selectfont
\begin{tabular}[t]{llcccccc}
\toprule
 &  & \multicolumn{3}{c}{ $ p = 200 $} & \multicolumn{3}{c}{ $ p = 300 $}\\
\midrule
Structures & Measures & SBM & STS & AT & SBM & STS & AT\\
\cmidrule{1-8}
 &  & \textbf{0.0175} & 0.0384 & 0.0225 & \textbf{0.0186} & 0.0312 & 0.0216\\

 & \multirow[t]{-2}{*}{\raggedright\arraybackslash \texttt{rmse}} & (0.0007) & (0.0002) & (0.0011) & (0.0006) & (0.0002) & (0.0007)\\

 &  & \textbf{0.5289} & 0.7292 & 0.6164 & \textbf{0.6779} & 0.7662 & 0.7383\\

\multirow[t]{-4}{*}{\raggedright\arraybackslash \textbf{Random}} & \multirow[t]{-2}{*}{\raggedright\arraybackslash \texttt{mnorm}} & (0.0554) & (0.0125) & (0.0333) & (0.1219) & (0.0167) & (0.0368)\\
\cmidrule{1-8}
 &  & \textbf{0.0105} & 0.0155 & 0.0109 & \textbf{0.0083} & 0.0111 & 0.0089\\

 & \multirow[t]{-2}{*}{\raggedright\arraybackslash \texttt{rmse}} & (0.0006) & (0.0003) & (0.0008) & (0.0003) & (0.0002) & (0.0007)\\

 &  & \textbf{0.3683} & 0.4833 & 0.4065 & \textbf{0.3927} & 0.4922 & 0.4417\\

\multirow[t]{-4}{*}{\raggedright\arraybackslash \textbf{Hubs}} & \multirow[t]{-2}{*}{\raggedright\arraybackslash \texttt{mnorm}} & (0.0484) & (0.006) & (0.0491) & (0.0505) & (0.0082) & (0.0497)\\
\cmidrule{1-8}
 &  & \textbf{0.0107} & 0.0132 & 0.0116 & \textbf{0.0085} & 0.0105 & 0.0091\\

 & \multirow[t]{-2}{*}{\raggedright\arraybackslash \texttt{rmse}} & (0.0006) & (0.0003) & (0.0007) & (0.0004) & (0.0002) & (0.0007)\\

 &  & \textbf{0.3709} & 0.4362 & 0.4051 & \textbf{0.4018} & 0.4412 & 0.4261\\

\multirow[t]{-4}{*}{\raggedright\arraybackslash \textbf{Cliques}} & \multirow[t]{-2}{*}{\raggedright\arraybackslash \texttt{mnorm}} & (0.0588) & (0.0181) & (0.0396) & (0.0527) & (0.0209) & (0.0328)\\
\bottomrule
\end{tabular}
\end{table}

Next, we consider high-dimensional settings, where $p\in \{200, 300\}$.
Due to their prohibitively slow calculation speed and repeated numerical errors, the two Bayesian approaches, SSSL and BM prior, failed to infer in this scenario, while the SBM prior completes inference without producing a numerical error. 
For that reason, we compare the SBM prior only with the frequentist methods, STS and AT estimators.
Note that this observation demonstrates the relative benefits of the proposed SBM prior.
In Table \ref{tab:result2}, we summarize the averages and standard deviations (in parentheses) for the high-dimensional settings, where $p\in \{200, 300\}$.
In terms of {\tt rmse}, the SBM prior outperforms other methods, while the AT estimator shows the best performance in terms of {\tt mnorm}.
Thus, a similar phenomenon to the previous $p=100$ scenario can be seen.

\subsection{Varying $n$ and $p$}\label{sec:simulation2}
In this subsection, we examine scenarios where both the sample size and dimension vary simultaneously. 
First, we set the range of dimensions as $p \in \{50, 150, 300, 500\}$.
For each dimension $p$, we determine the sample size $n$ according to the following two scenarios: (1) the sample size is equal to the dimension ($n=p$) and (2) the sample size is twice the dimension ($n = 2p$). 

\begin{figure}[!ht]
	\centering
    \begin{subfigure}{0.95\textwidth}
    \centering
    \includegraphics[width=1\textwidth]{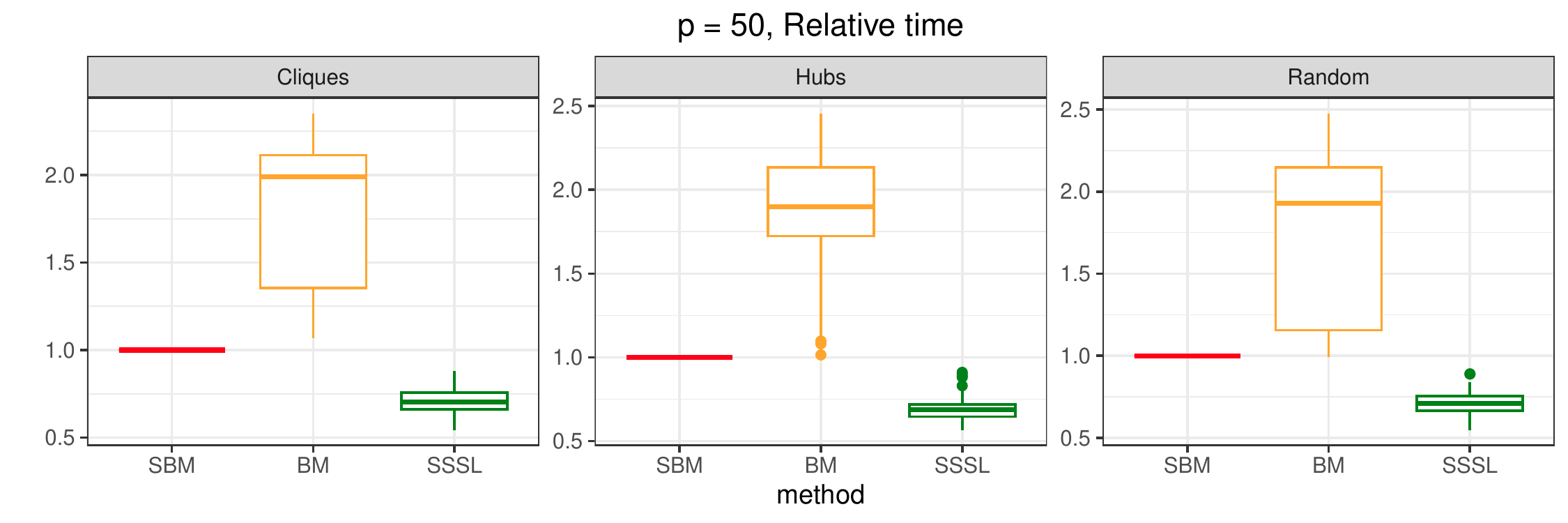}
    \end{subfigure}
    \begin{subfigure}{0.95\textwidth}
    \centering
    \includegraphics[width=1\textwidth]{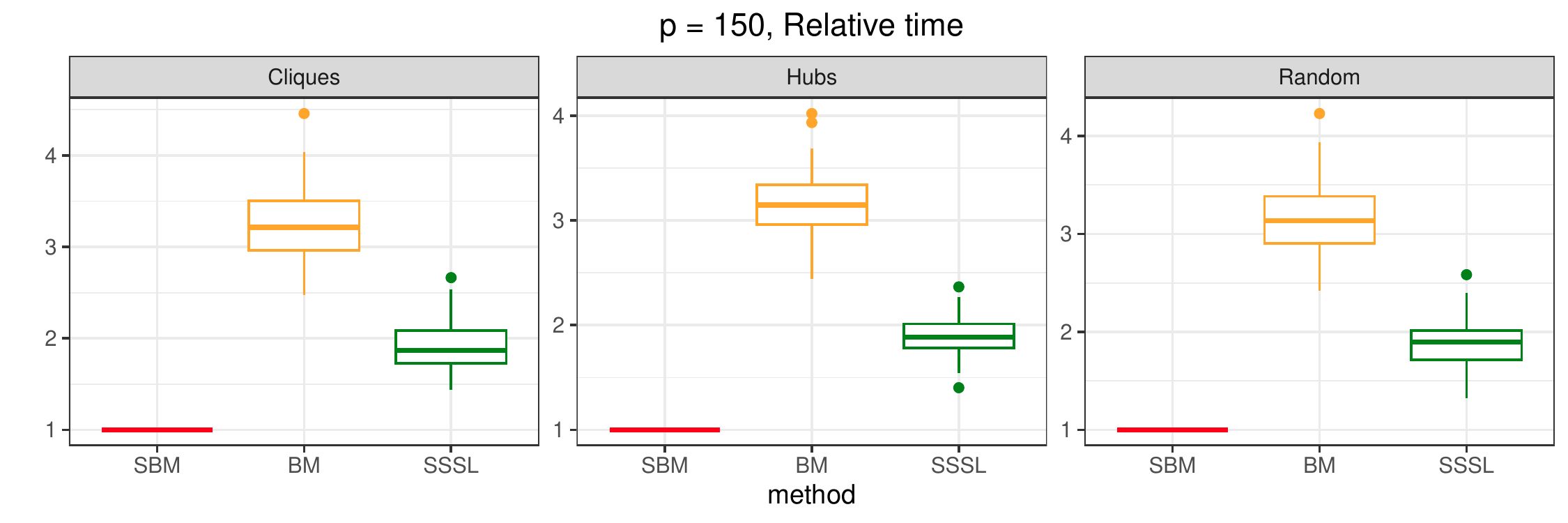}
    \end{subfigure}
    \begin{subfigure}{0.95\textwidth}
    \centering
    \includegraphics[width=1\textwidth]{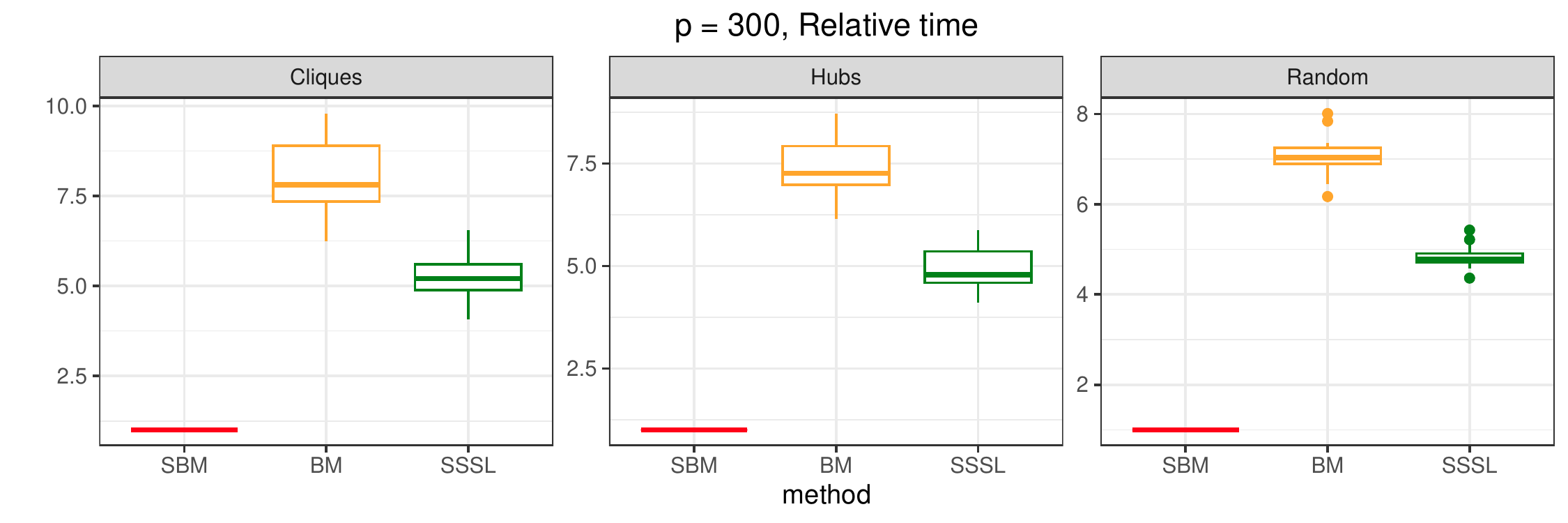}
    \end{subfigure}
	\caption{Boxplot of the relative computing times of Bayesian methods for $n = 2p$, where the dimension $p$ varies with $p\in \{50,150,300\}$.}
	\label{fg:reltime}
\end{figure}

To demonstrate the computational efficiency of the proposed SBM prior, we present boxplots illustrating the relative computing times for Bayesian methods, SBM, SSSL, and BM, in Figure \ref{fg:reltime}. 
The relative computing time is calculated by dividing the computing time of other Bayesian methods by that of the SBM prior.
This figure illustrates that the proposed approach is faster than the BM prior and the SSSL prior in high-dimensional settings. 
It suggests that the SBM prior dramatically improves the computational speed over existing Bayesian methods by incorporating the screening procedure as desired.

\begin{figure}[!ht]
\centering
    \begin{subfigure}{0.95\textwidth}
    \centering
    \includegraphics[width=0.8\textwidth]{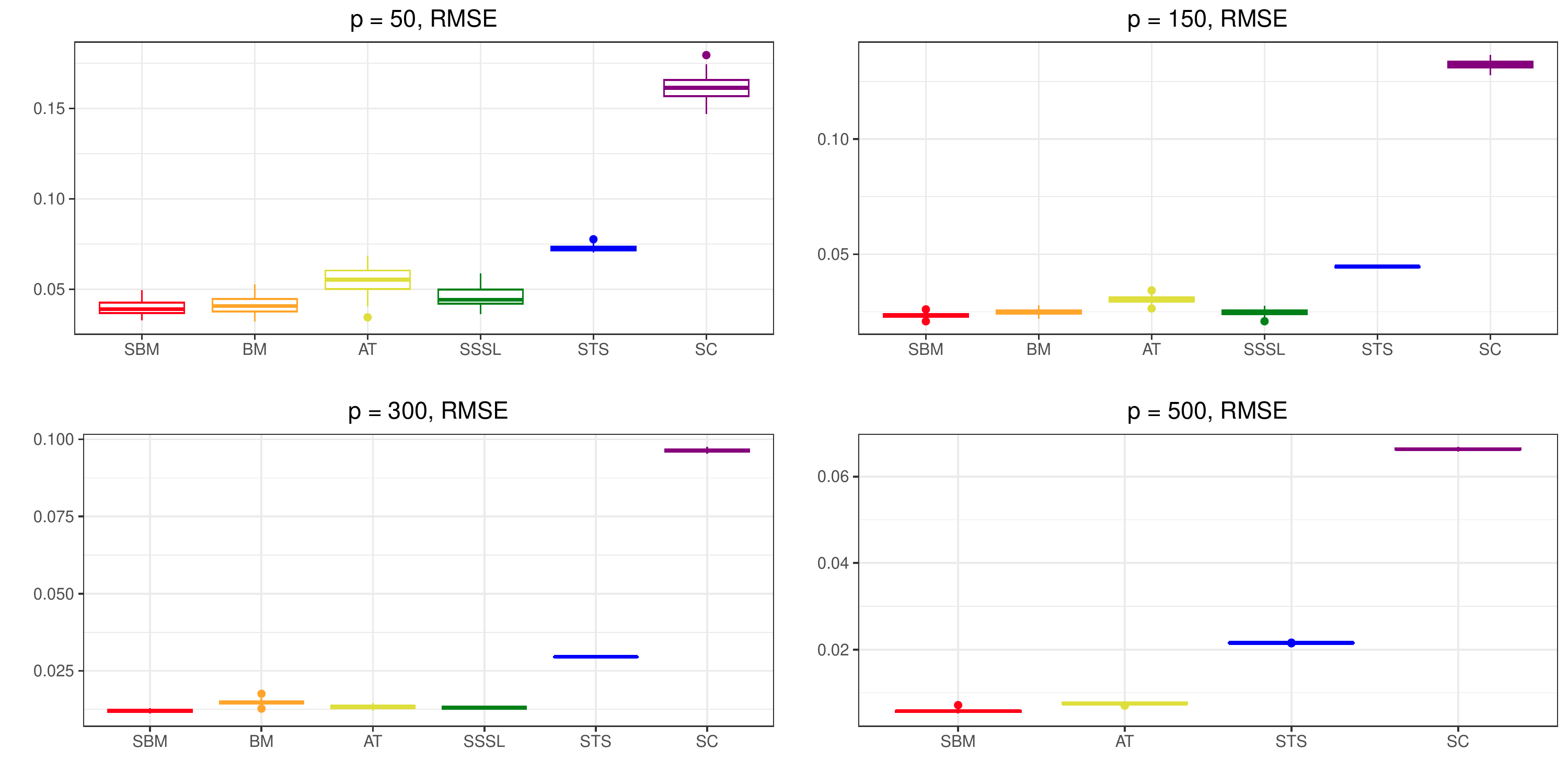}
    \caption{Random}
    \end{subfigure}
    \begin{subfigure}{0.95\textwidth}
    \centering
    \includegraphics[width=0.8\textwidth]{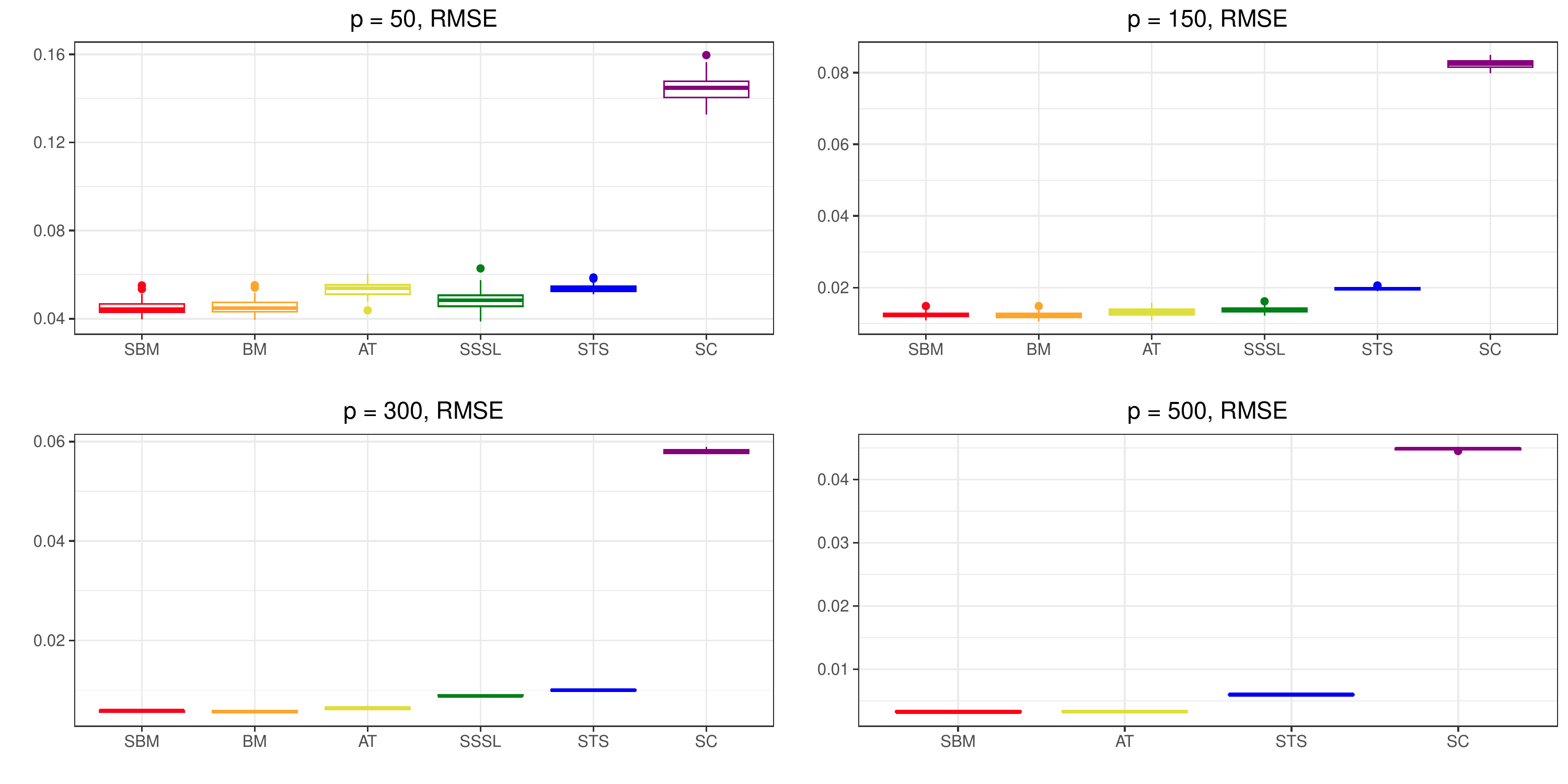}
    \caption{Hubs}
    \end{subfigure}
    \begin{subfigure}{0.95\textwidth}
    \centering
    \includegraphics[width=0.8\textwidth]{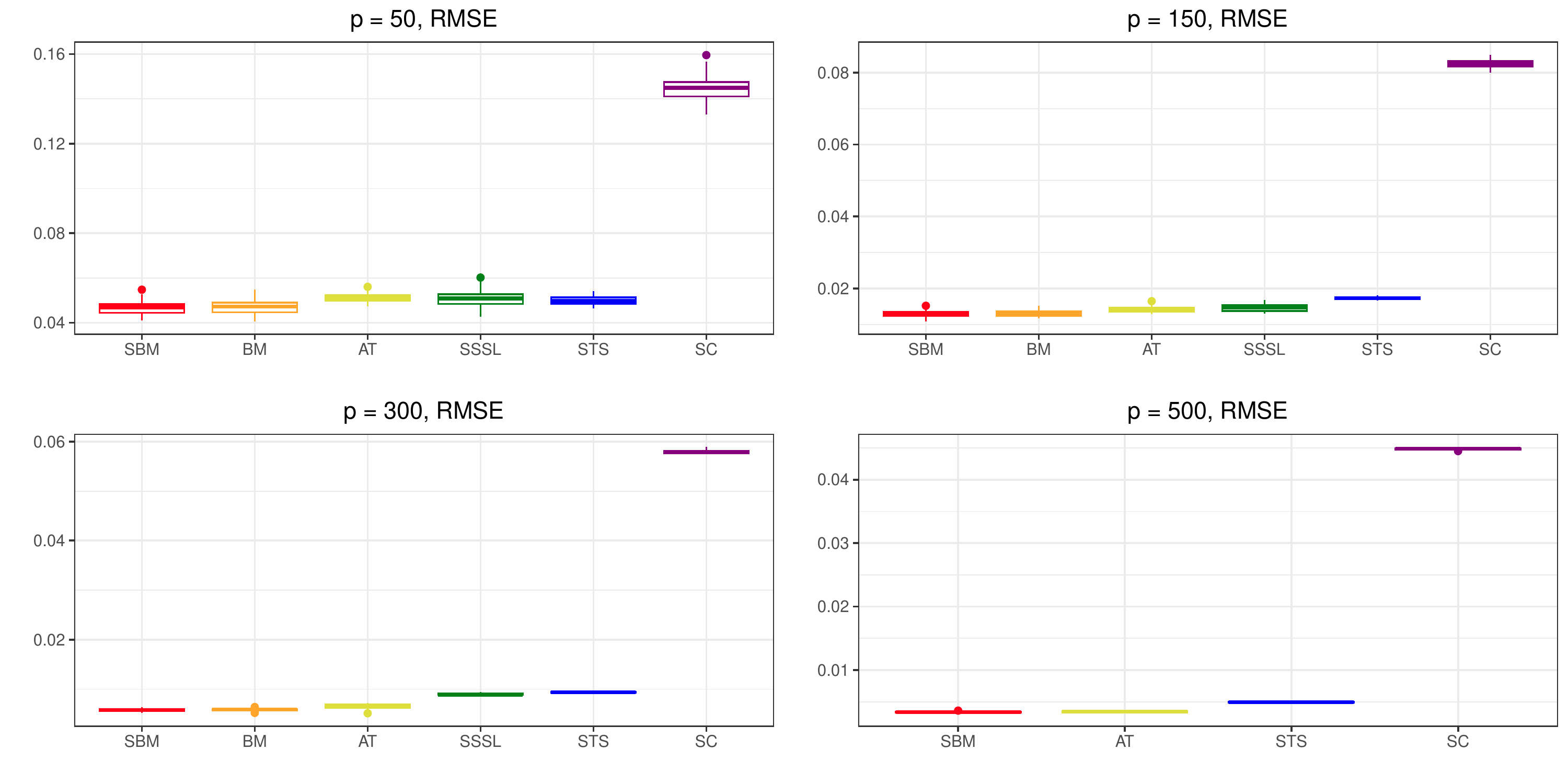}
    \caption{Cliques}
    \end{subfigure}
	\caption{Boxplot of the root mean squared errors (\texttt{rmse}) for $n = p$, where the dimension $p$ varies with $p\in \{50,150,300,500\}$.}
	\label{fg:rmse_boxplot1}
\end{figure}


\begin{figure}[!ht]
	\centering
    \begin{subfigure}{0.95\textwidth}
    \centering
    \includegraphics[width=0.8\textwidth]{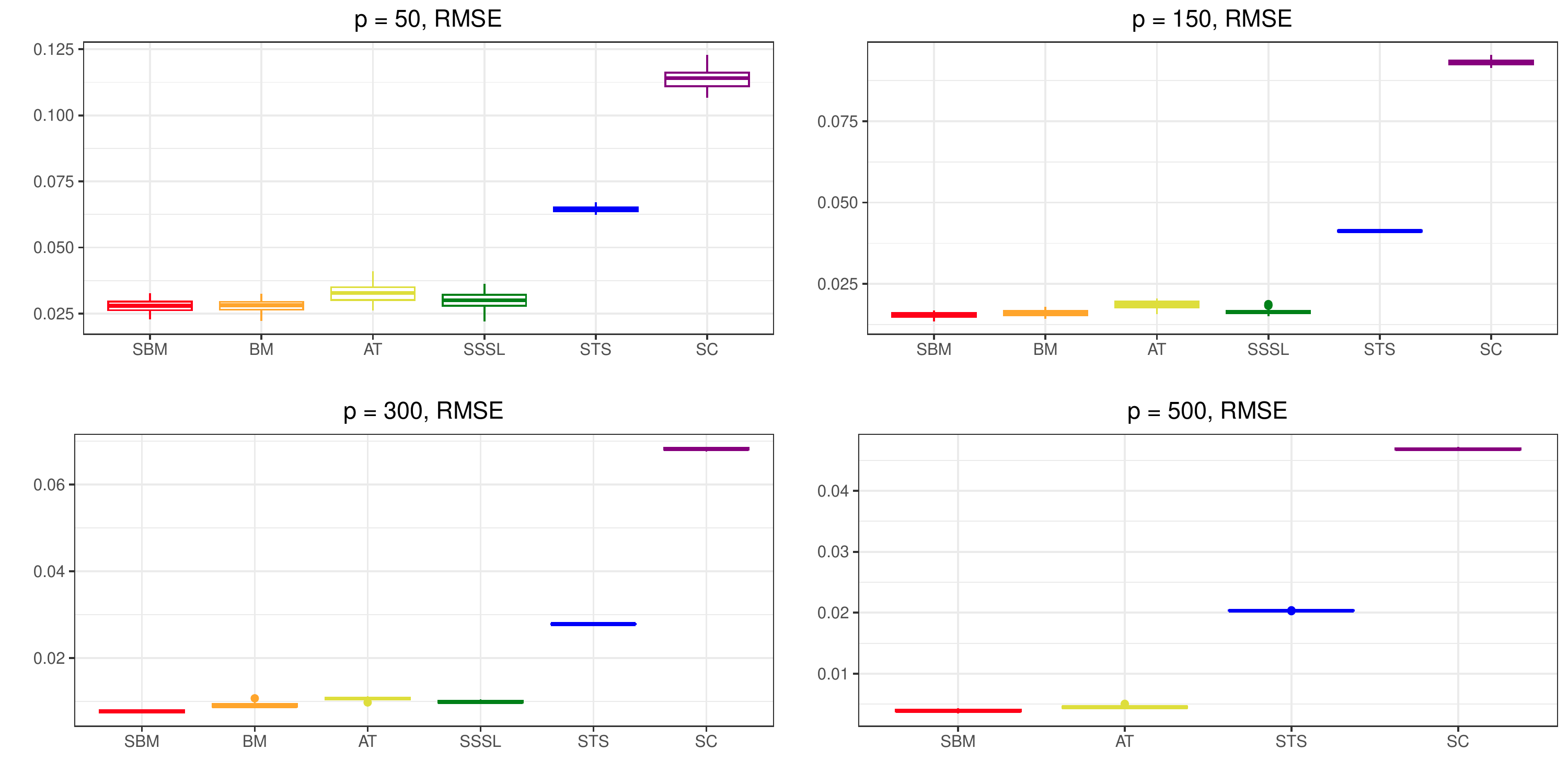}
    \caption{Random}
    \end{subfigure}
    \begin{subfigure}{0.95\textwidth}
    \centering
    \includegraphics[width=0.8\textwidth]{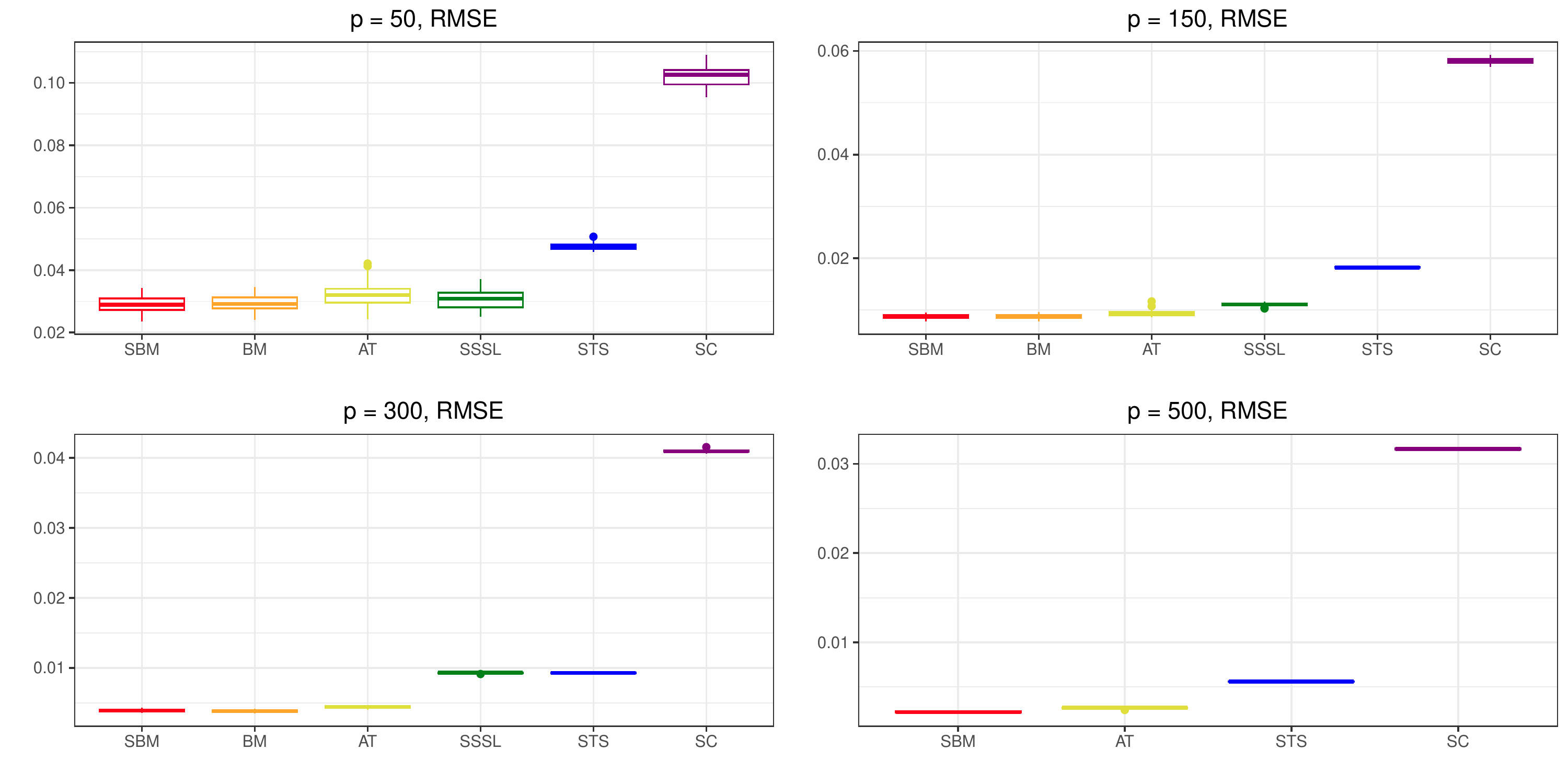}
    \caption{Hubs}
    \end{subfigure}
    \begin{subfigure}{0.95\textwidth}
    \centering
    \includegraphics[width=0.8\textwidth]{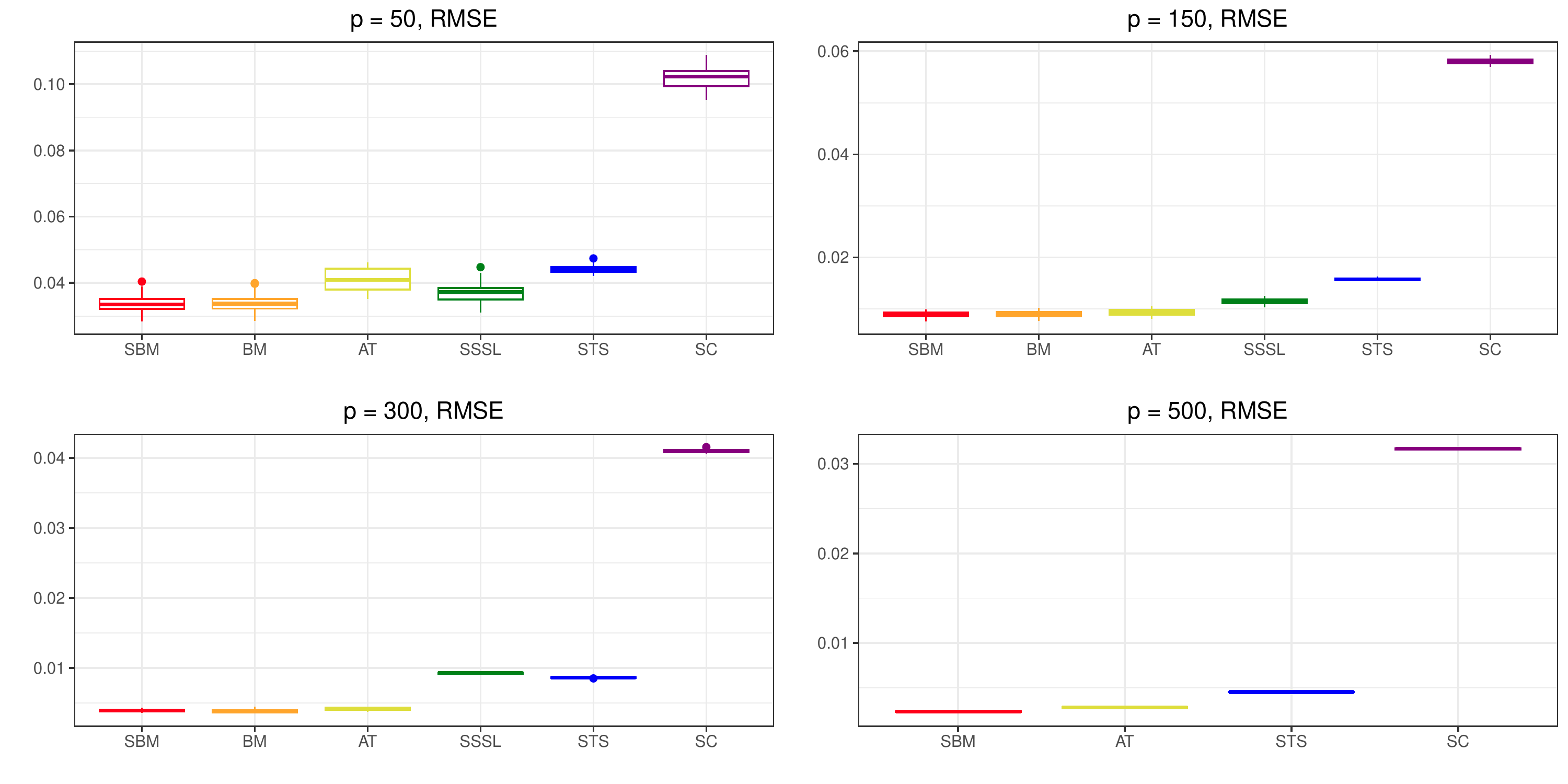}
    \caption{Cliques}
    \end{subfigure}
	\caption{Boxplot of the root mean squared errors (\texttt{rmse}) for $n = 2p$, where the dimension $p$ varies with $p\in \{50,150,300,500\}$.}
	\label{fg:rmse_boxplot2}
\end{figure}

\begin{figure}[!ht]
	\centering
    \begin{subfigure}{0.95\textwidth}
    \centering
    \includegraphics[width=0.8\textwidth]{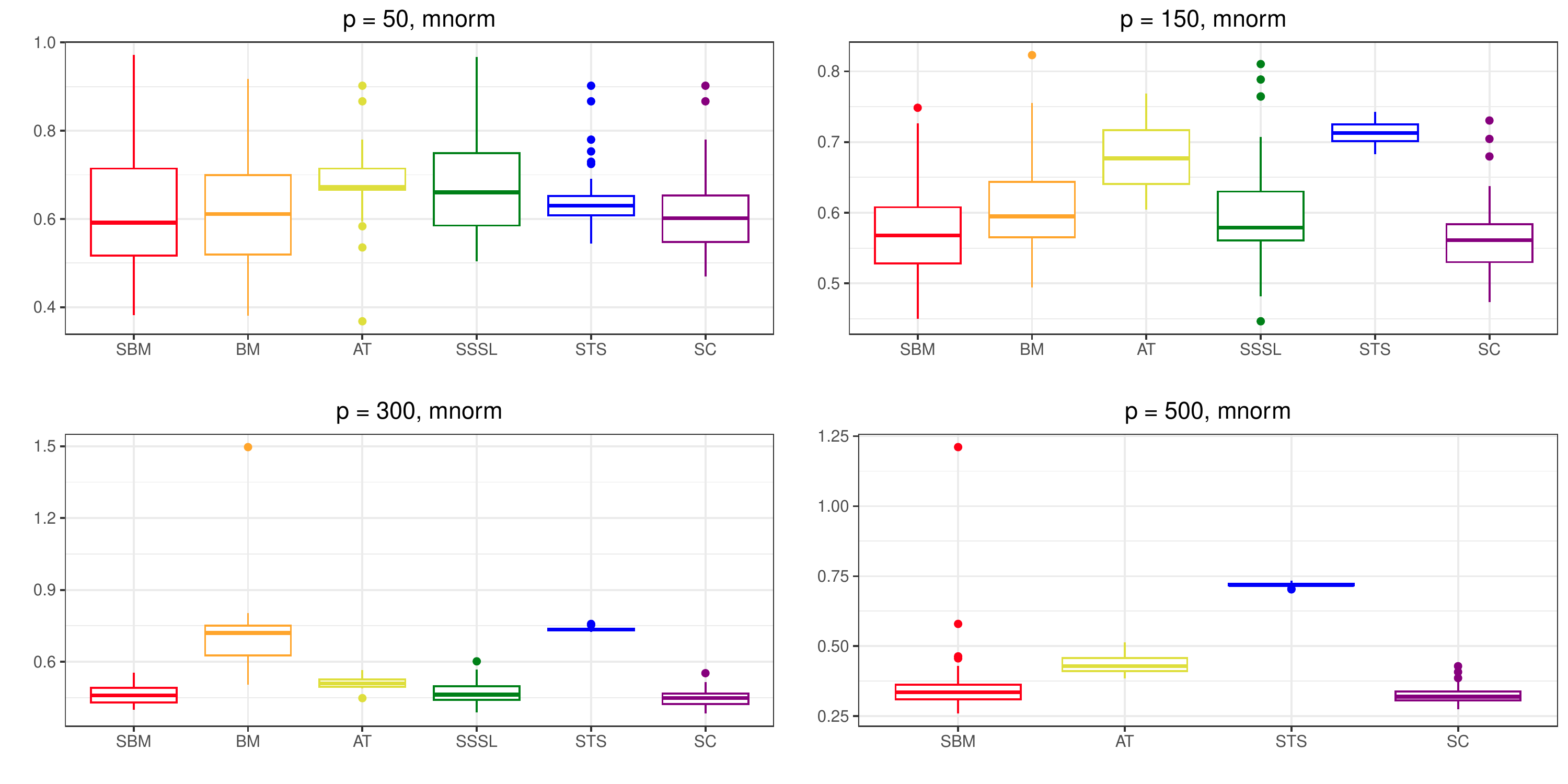}
    \caption{Random}
    \end{subfigure}
    \begin{subfigure}{0.95\textwidth}
    \centering
    \includegraphics[width=0.8\textwidth]{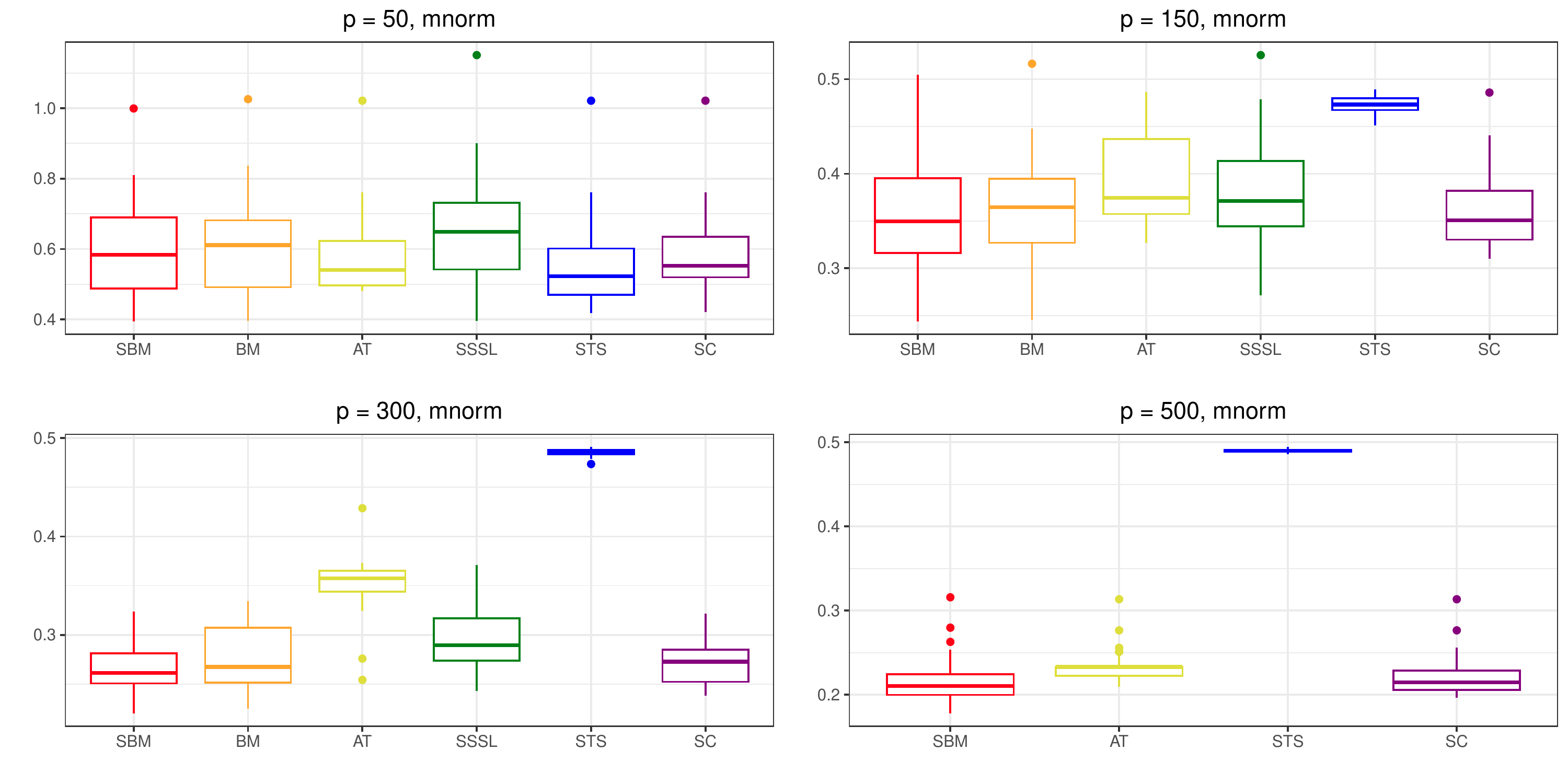}
    \caption{Hubs}
    \end{subfigure}
    \begin{subfigure}{0.95\textwidth}
    \centering
    \includegraphics[width=0.8\textwidth]{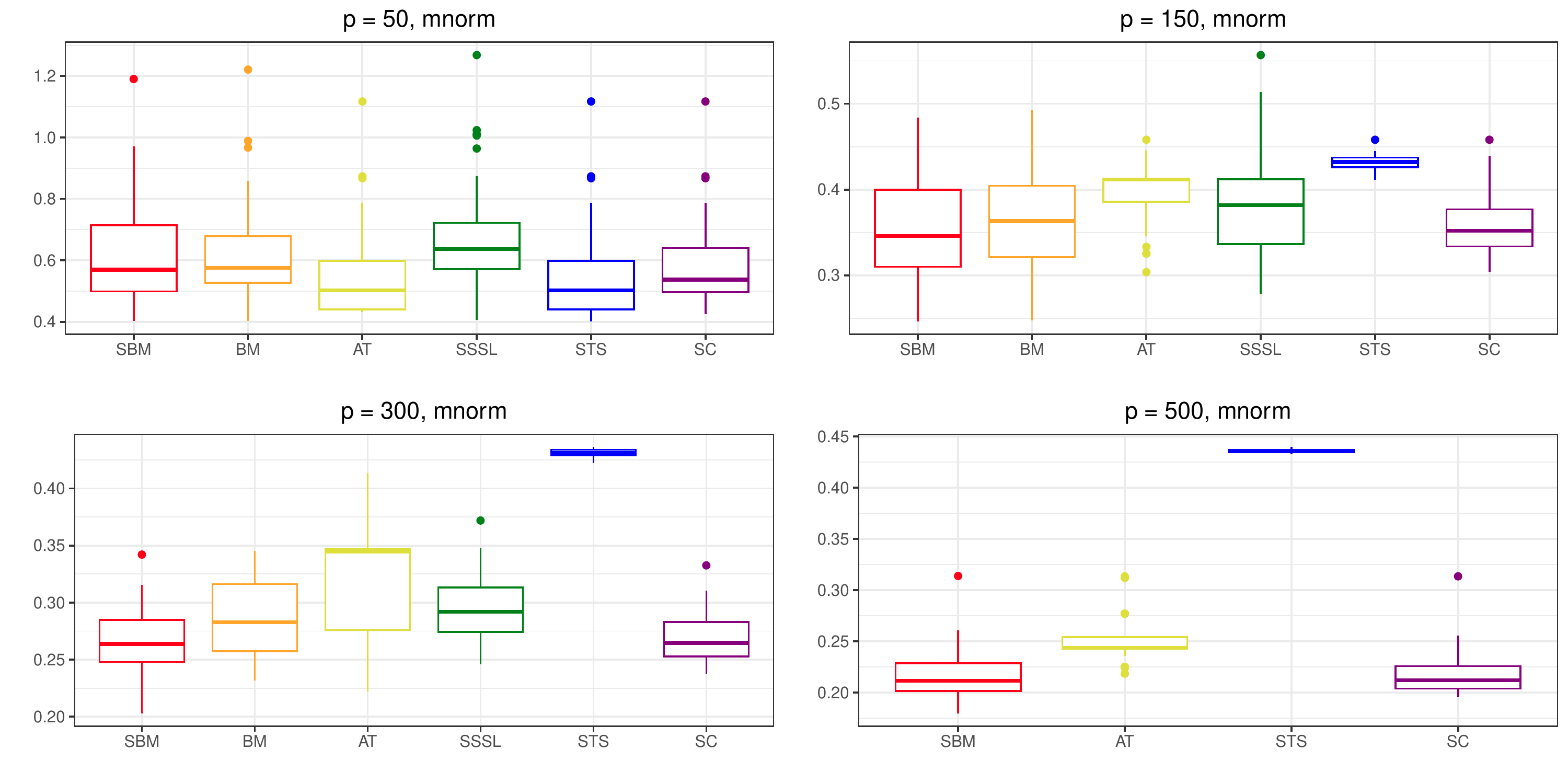}
    \caption{Cliques}
    \end{subfigure}
	\caption{Boxplot of the element-wise maximum norm (\texttt{mnorm}) values for $n = p$, where the dimension $p$ varies with $p\in \{50,150,300,500\}$.}
	\label{fg:mnorm_boxplot1}
\end{figure}

\begin{figure}[!ht]
	\centering
    \begin{subfigure}{0.95\textwidth}
    \centering
    \includegraphics[width=0.8\textwidth]{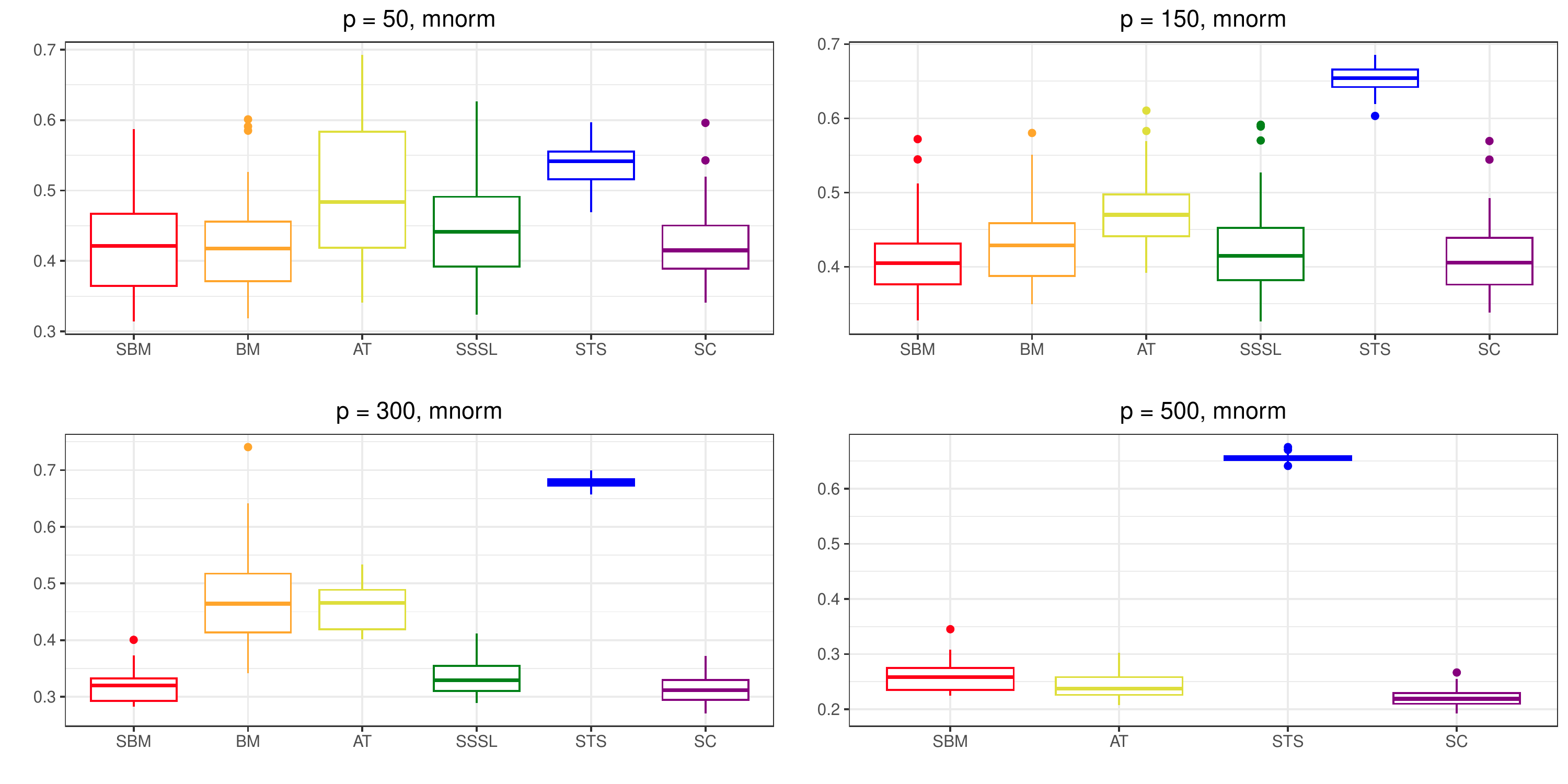}
    \caption{Random}
    \end{subfigure}
    \begin{subfigure}{0.95\textwidth}
    \centering
    \includegraphics[width=0.8\textwidth]{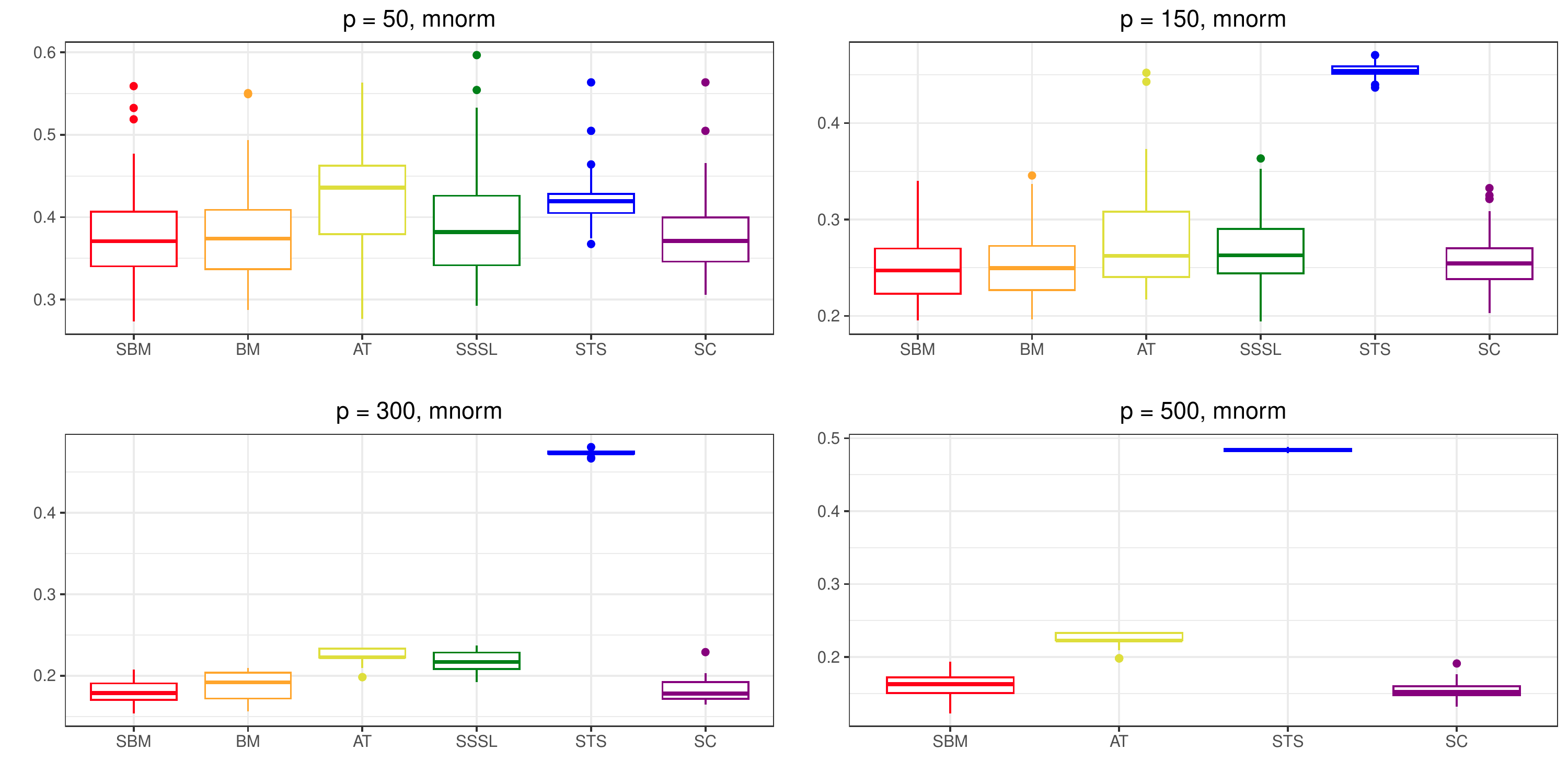}
    \caption{Hubs}
    \end{subfigure}
    \begin{subfigure}{0.95\textwidth}
    \centering
    \includegraphics[width=0.8\textwidth]{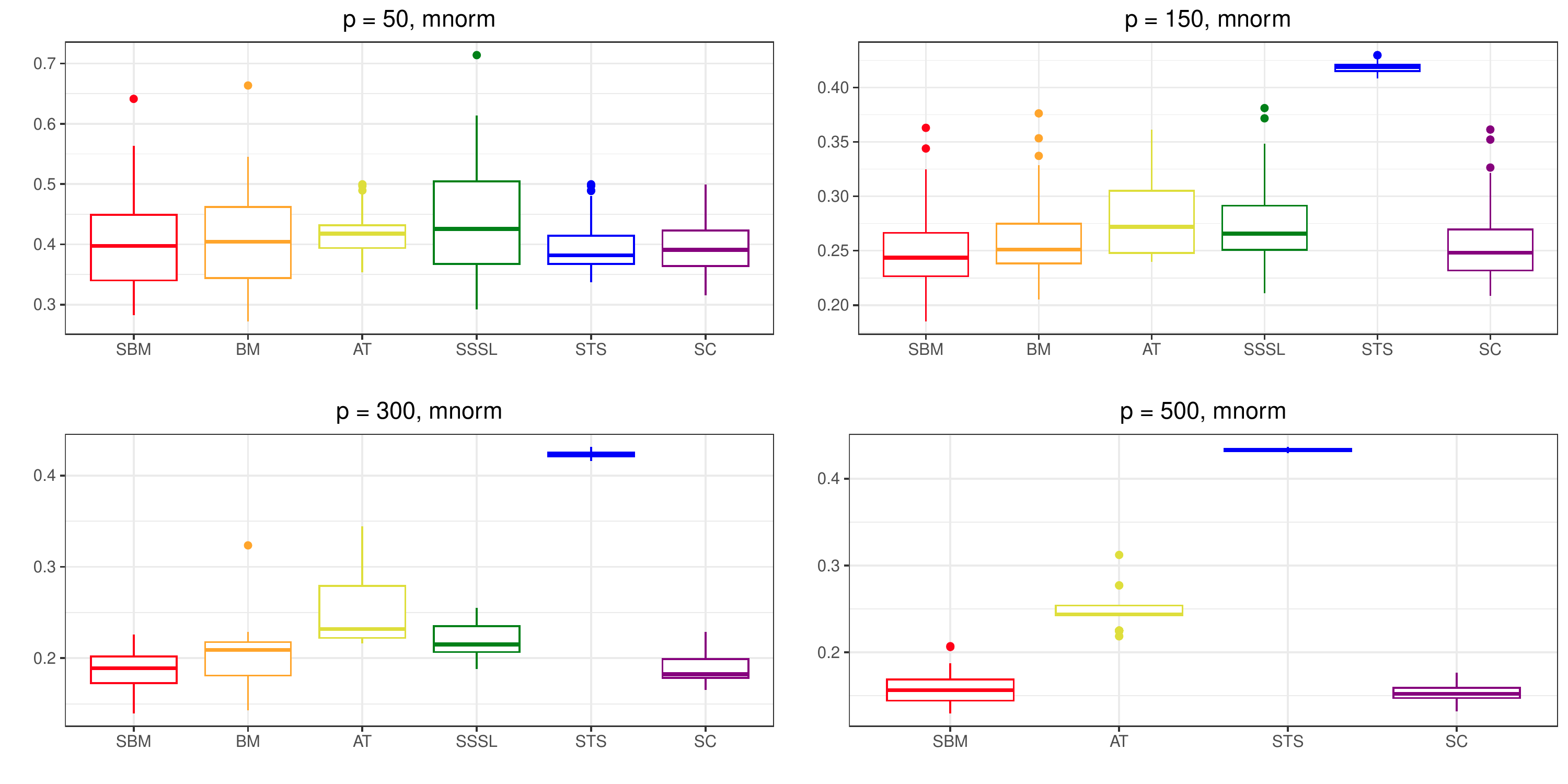}
    \caption{Cliques}
    \end{subfigure}
	\caption{Boxplot of the element-wise maximum norm (\texttt{mnorm}) values for $n = 2p$, where the dimension $p$ varies with $p\in \{50,150,300,500\}$.}
	\label{fg:mnorm_boxplot2}
\end{figure}

Figures \ref{fg:rmse_boxplot1} and \ref{fg:rmse_boxplot2} show the boxplots of the {\tt rmse} for the case where $n = p$ and $n=2p$, respectively.
When $p = 500$, BM and SSSL priors failed to infer due to the numerical error and memory issues involved in large-dimensional matrix operations, so we omit their boxplots for $p=500$.
Overall, in terms of the {\tt rmse}, the proposed SBM prior consistently achieves the best performance in most settings. 
It supports the theoretical result that the SBM attains the minimax optimal posterior convergence rate under the Frobenius norm.
Although the BM prior has similar performances to the SBM, the SBM exhibits significantly faster computation time.
Another Bayesian competitor, the SSSL method, yields slightly higher {\tt rmse} values compared to the SBM and also requires considerably longer computation time.
The AT estimator of course leads to much faster inference compared with the Bayesian methods, but it gives relatively large {\tt rmse} values in most settings. 
The SC and STS methods are outperformed by other methods, which implies that these methods are not appropriate for high-dimensional settings.
As the sample size grows from $n=p$ (Figure \ref{fg:rmse_boxplot1}) to $n=2p$ (Figure \ref{fg:rmse_boxplot2}), the performances of all methods tend to improve while maintaining the relative performance of each method.

Figures \ref{fg:mnorm_boxplot1} and \ref{fg:mnorm_boxplot2} show the boxplots of the {\tt mnorm} for the case where $n = p$ and $n=2p$, respectively.
For the same reason as previously described in the {\tt rmse} case, the results of BM and SSSL are omitted when $p = 500$.
The SBM prior performs reasonably well and outperforms other contenders, except the SC, in most settings. 
Note that when evaluating performance based on the element-wise maximum norm, it is not surprising that the SC shows better performance compared to other methods specifically designed for high-dimensional settings.
Other Bayesian methods, the BM prior and SSSL, tend to have higher {\tt mnorm} than the SBM prior.
Similar to the {\tt rmse} case, the AT estimator obtains relatively large {\tt mnorm} values compared to other high-dimensional methods.

\section{Real Data Application}\label{sec:realdata}
We carry out the linear discriminant analysis (LDA) classification \citep{anderson2003introduction}
to illustrate the performance of the proposed SBM prior by applying it to two high-dimensional gene datasets: (1) colon cancer data introduced in \cite{alon1999} and (2) small round blue-cell tumors (SRBCT) data described in \cite{khan2001classification}. 

\begin{figure}[!tb]
    \centering
    \begin{subfigure}{0.95\textwidth}
    \centering
    \includegraphics[width=\textwidth]{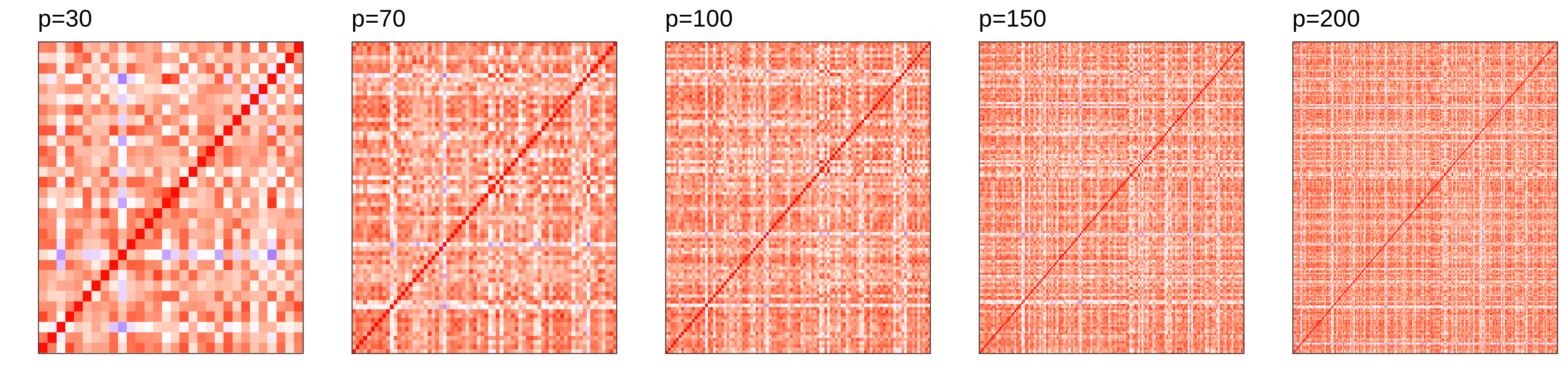}
    \caption{colon cancer dataset}
    \end{subfigure}
    \begin{subfigure}{0.95\textwidth}
    \centering
    \includegraphics[width=\textwidth]{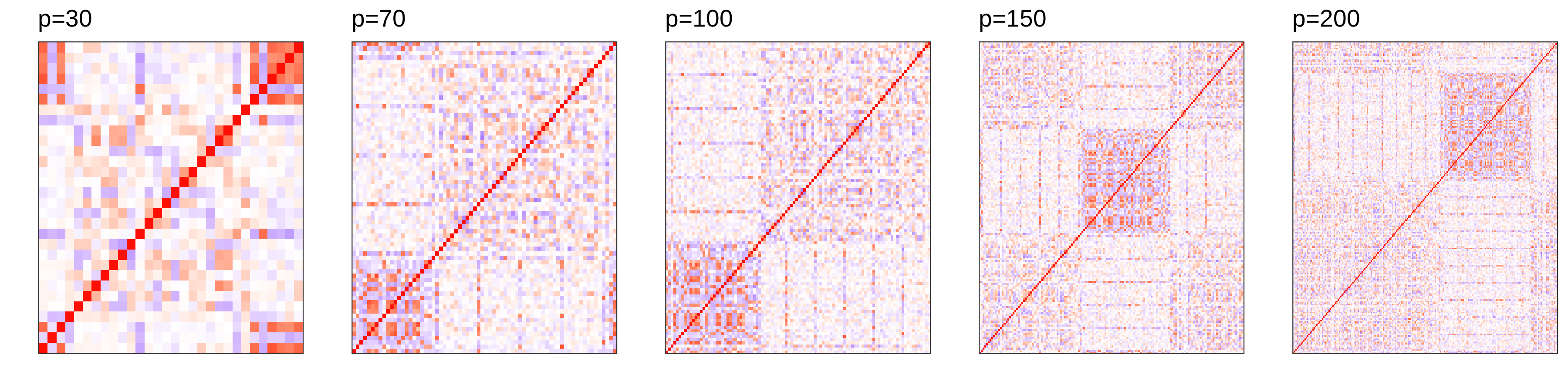}
    \caption{SRBCT dataset}
    \end{subfigure}
    \caption{Sample correlation matrices of colon cancer and SRBCT datasets.}
	\label{fg:realdata-cor}
\end{figure}

The colon cancer dataset is available at \url{http://genomics-pubs.princeton.edu/oncology/affydata/}, and it contains 2,000 gene expression values from 22 non-tumor tissues and 40 tumor tissues $(n=62)$.
Following the approach of \cite{rothman08}, we select $30,~70,~100,~150,$ and 200 most significant genes based on the two-sample $t$-statistic.
The SRBCT dataset is available at {\tt plsgenomics} package in \textsf{R} from the microarray experiments of Small Round Blue Cell Tumors (SRBCT) of childhood cancer study. It contains 2,308 gene expression values from 83 samples $(n=83)$, which consist of 29, 11, 18 and 25 tissue samples from the four types of tumors (Ewing sarcoma, Burkitt lymphoma, neuroblastoma and rhabdomyosarcoma).
To reduce the number of genes, similar to \cite{rothman2009generalized} and \cite{cai2011adaptive}, we select top $0.4p$ and bottom $0.6p$ genes for $p=30,~70,~100,~150,$ and 200 based on $F$-statistic.
Figure \ref{fg:realdata-cor} visualizes the sample correlation matrices of the two datasets for each value of $p$. They both show a distinctive sparse structure.
Note that for both datasets, BM and SSSL produced numerical errors and failed to perform possibly due to relatively large $p$ compared with the sample size $n$.

We apply LDA to the two datasets for classifying each observation $x$ based on the following rule:
\begin{align*}
	\delta_k(x) =& \argmax_{1\le k \le K} \Big( x^T \hat{\Sigma}^{-1} \hat{\mu}_k - \frac{1}{2} \hat{\mu}_k^T \hat{\Sigma}^{-1} \hat{\mu}_k + \log \hat{\pi}_k   \Big), \quad k \in \{1,\ldots, K\}, 
\end{align*}
where $\hat{\mu}_k$ and $\hat{\pi}_k$ are the sample mean and proportion, respectively, for class $k \in \{1,\ldots, K\}$.
The estimated covariance matrix $\hat{\Sigma}$ is obtained by using SBM prior, STS and AT estimators.
Note that we have $K=2$ and $K=4$ for the colon cancer and SRBCT datasets, respectively.


\begin{figure}[!tb]
    \centering
    \begin{subfigure}{0.48\textwidth}
    \centering
    \includegraphics[width=\textwidth]{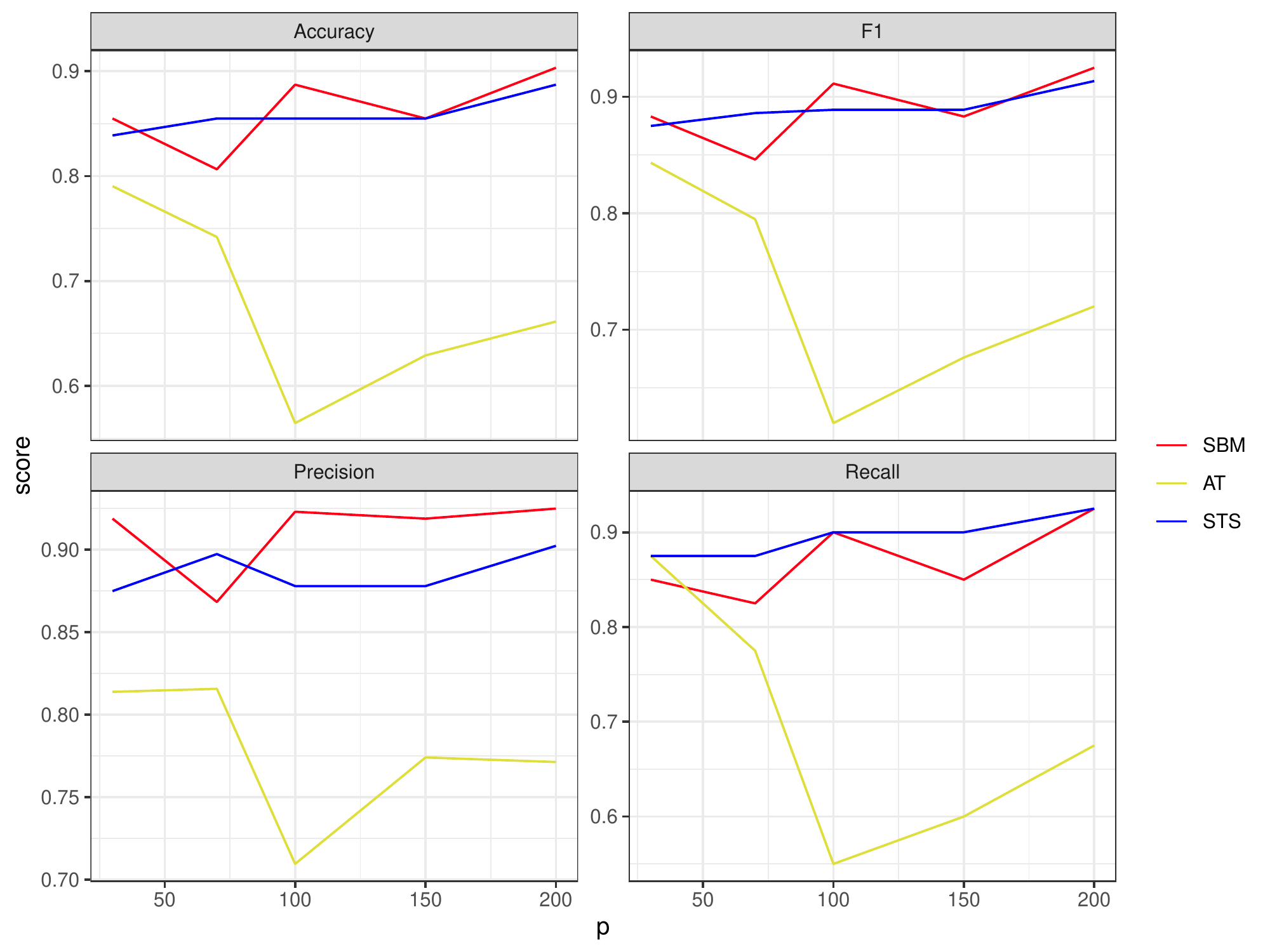}
    \caption{colon cancer dataset}
    \end{subfigure}
    \begin{subfigure}{0.48\textwidth}
    \centering
    \includegraphics[width=\textwidth]{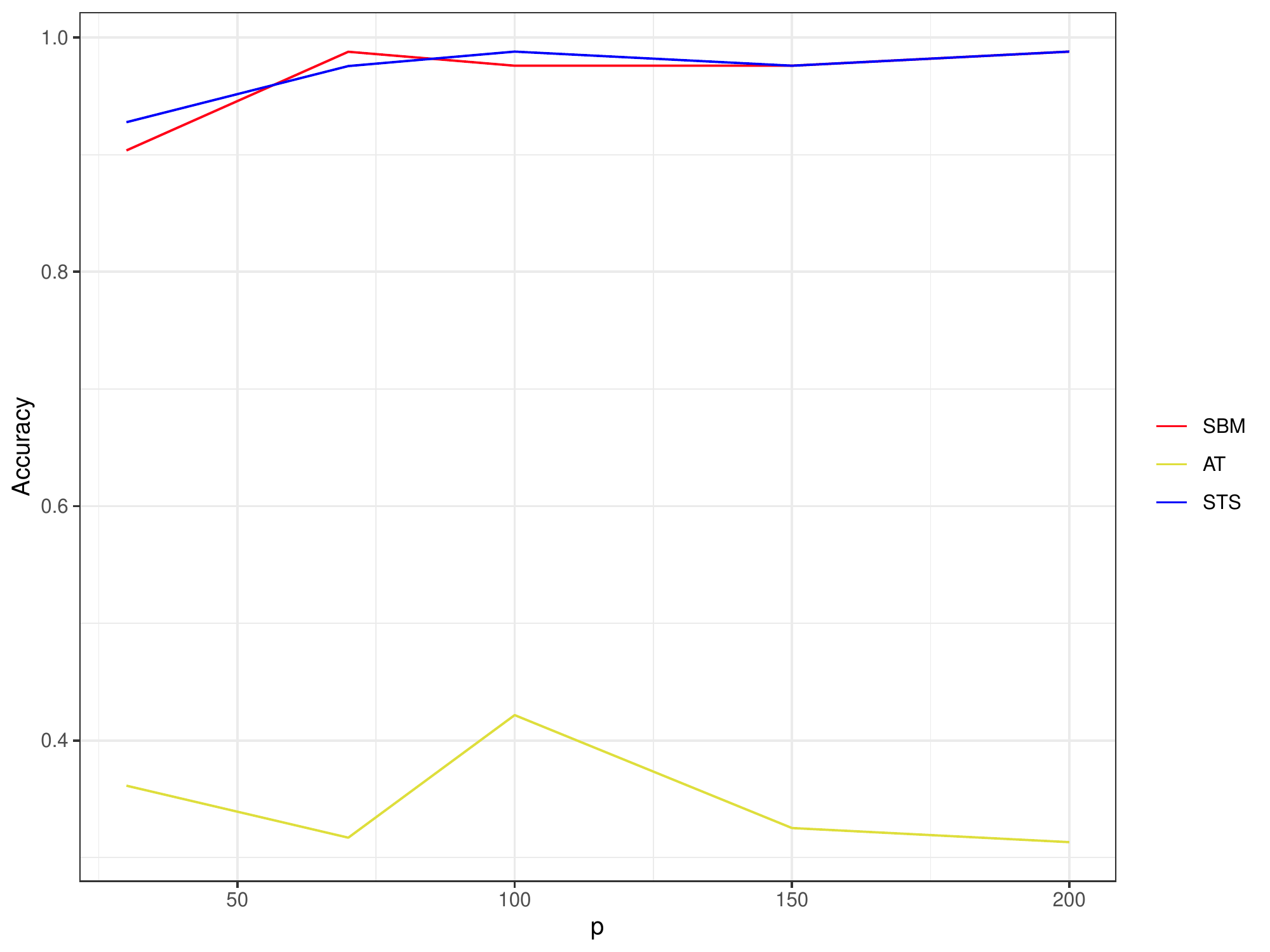}
    \caption{SRBCT dataset}
    \end{subfigure}
    \caption{Classification scores for colon cancer and SRBCT datasets.}
	\label{fg:realdata}
\end{figure}

The results in Figure \ref{fg:realdata}
show the classification scores (accuracy, precision, recall and F1 scores) obtained from the leave-one-out cross-validation (LOOCV) for each method. 
For both datasets, the SBM prior shows comparable performance to STS and outperforms AT.
In particular, in the case of the SRBCT data, the SBM prior performed well even though the number of samples $n$ is relatively small compared with the dimension $p$.
These results suggest that the proposed method can be used for high-dimensional data that existing Bayesian methods cannot handle, and that it performs well in applications such as LDA.

\section{Discussion}\label{sec:disc}
In this paper, we propose the SBM shrinkage prior for high-dimensional sparse covariance matrix estimation. 
The SBM shrinkage prior imposes beta-mixture shrinkage priors on the off-diagonal entries selected by the sure screening method and sets the remaining off-diagonal entries to zero, under the positive definiteness constraint.
To choose a threshold for the screening operator, we suggest an FNR-based approach using Jeffreys’ default Bayes factor for correlation.
For theoretical justifications, we show that the screening procedure has sure screening property and selection consistency, and prove the posterior convergence rate of the SBM prior under the Frobenius norm. 
Numerical experiments show that the proposed Bayesian method can be applied to high-dimensional data in which existing Bayesian methods may collapse and can outperform other frequentist covariance estimators in terms of accurate covariance matrix estimation and lower misclassification error based on LDA.





\appendix
\section*{Appendix A. Proof of main results}\label{app:proof_main}



In this appendix, we present the proofs of Theorems \ref{thm:sure_scr}, \ref{thm:selection_cons} and \ref{thm:conv_rate} of the paper.

\noindent
{\bf Proof of Theorem \ref{thm:sure_scr}}. 
Note that 
	\bea
	\bbP_0 (S_0(\sg_0)  \nsubseteq S_r(\hat{R}) ) 
	&\le& \bbP_0 \Big( \cup_{(i,j)\in S_0(\sg_0)} \{(i,j) \notin S_r(\hat{R}) \} \Big) \\
	&\le& \sum_{(i,j)\in S_0(\sg_0)} \bbP_0 (|\hat{\rho}_{ij}| \le r)  \\
	&\le& \sum_{(i,j)\in S_0(\sg_0)} \frac{23 \, ( n \rho_{0,ij} r \wedge 1)}{\sqrt{n} ( |\rho_{0,ij}| - r) }  \exp \Big\{  - \frac{n (|\rho_{0,ij}|-r)^2  }{2 (1+ \rho_{0,ij}^2 )}    \Big\} \\
	&\le& \sum_{(i,j)\in S_0(\sg_0)} \frac{23 \, ( n  r \wedge 1)}{\sqrt{n} ( |\rho_{0,ij}| - r) }  \exp \Big\{  - \frac{n (|\rho_{0,ij}|-r)^2  }{2 (1+ \rho_{0,ij}^2 )}    \Big\}  \\
	&\le& \frac{23 s_0 ( n  r \wedge 1)}{ (2+\delta) \sqrt{\log (n\vee p)} }  \exp \Big\{  - \frac{(2+\delta)^2}{2} \log (n\vee p)   \Big\} 
	\eea
	by Lemma \ref{lem:samp_corr_small_prob} and condition (A1). \hfill $\blacksquare$

\noindent
{\bf Proof of Theorem \ref{thm:selection_cons}}. Note that 
	\bea
	&& \bbP_0( S_0(\sg_0) \neq S_r(\hat{R}))  \\
	&=& \bbP_0 ( S_0(\sg_0) \nsubseteq S_r(\hat{R})) + \bbP_0 ( S_0(\sg_0) \subsetneq S_r(\hat{R})) \\
	&\le& \bbP_0 ( S_0(\sg_0) \nsubseteq S_r(\hat{R})) +  \sum_{(i,j)\notin S_0(\sg_0)} \bbP_0 (|\hat{\rho}_{ij}| \ge r ) \\
	&\le& \bbP_0 ( S_0(\sg_0) \nsubseteq S_r(\hat{R})) +  \frac{4\{p(p-1)/2 - s_0 \} }{r \sqrt{n} } \exp \Big( - \frac{n}{2} r^2 \Big)  \\
	&\le& \frac{23 s_0 ( n  r \wedge 1)}{ (2+\delta) \sqrt{\log (n\vee p)} }  \exp \Big\{  - \frac{(2+\delta)^2}{2} \log (n\vee p)   \Big\}  
	+ \frac{2\{p(p-1) - 2s_0 \} }{C_{\rm th} \sqrt{\log (n\vee p)} } (n\vee p)^{-\frac{C_{\rm th}^2}{2} }
	\eea
	by Lemmas \ref{lem:samp_corr_concen} and \ref{lem:samp_corr_small_prob}.
	The last display is of order $o(1)$ when $C_{\rm th} \ge 2$. \hfill $\blacksquare$

\noindent
{\bf Proof of Theorem \ref{thm:conv_rate}}. We closely follow the proof of Theorem 3.1 in \cite{lee2021betamixture}.
	Let $\epsilon_n = \sqrt{(p+s_0) \log p / n}$.
	With a slight abuse of notation, $\pi_r$ also denotes the prior for $f_\Sigma$ induced by the SBM prior, where $f_\sg$ is the probability density function of $N_p(0, \sg)$ based on $n$ random samples $X_1,\ldots, X_n$.
	By Theorem 2.1 in \cite{ghosal2000convergence} and Lemma 5.1 in \cite{lee2021betamixture}, it suffices to show that 
	\bean\label{suff_conditions}
	\log D(\epsilon_n, \calP_n ,  d)  
	&\le& C_1 n \epsilon_n^2 , \label{entropy} \\
	\pi_r ( \calP_n^c ) &\le& \exp \big\{  -(C_2 + 4) n \epsilon_n^2   \big\} , \label{sieve_mass} \\
	\pi_r( B_{\epsilon_n} ) &\ge& \exp \big( - C_2 n \epsilon_n^2 \big) , \label{prior_thickness}
	\eean
	for some sieve $\calP_n \subset \calP = \{ f_\sg : \sg \in \calC_p \}$ and constants $C_1, C_2>0$,
	where $d$ is the Hellinger metric, and $D(\epsilon_n, {\calP}_n ,  d)$ is the $\epsilon_n$-packing number of $\calP_n$ based on $d$,
	\bea
	B_\epsilon &:=& \Big\{ f_{\Sigma}: \,\, \Sigma \in \calC_p, \,\,  K( f_{\sg_0}, f_\Sigma )    < \epsilon^2, \,\,  V(f_{\sg_0} , f_\Sigma )  < \epsilon^2 \Big\}  , \\
	K( f_{\sg_0}, f_\Sigma )  & := & \int f_{\sg_0} (x) \log \frac{ f_{\sg_0} (x) }{ f_{\Sigma} (x) } dx ,  \\
	V(f_{\sg_0} , f_\Sigma )  & := & \int f_{\sg_0}  (x) \left( \log \frac{ f_{\sg_0} (x) }{ f_{\Sigma} (x) } \right)^2 dx .
	\eea
	
	For some constants $c_1>1$ and $c_2>0$, define the sieve 
	\bea
	\calP_n = \big\{ f_{\sg}: |S_{\delta_n}(\sg)|\le s_n, \, \sg\in \calU(\epsilon), \, \|\sg\|_{\max} \le L_n \big\},
	\eea
	where $\delta_n  = \epsilon^3 \epsilon_n$, $s_n = c_1 n \epsilon_n^2 / \log p$, $\|\sg\|_{\max} = \max_{j,k}|\sigma_{jk}|$ and $L_n = c_2n \epsilon_n^2$.
	By Theorem 5.2 of \cite{lee2021betamixture}, 
	\bea
	\log D(\epsilon_n, \calP_n ,  d)  
	&\le& \big( 12 + \beta^{-1} \big) c_1 n \epsilon_n^2 ,
	\eea
	thus we have proved \eqref{entropy} by taking $ (12 + \beta^{-1}) c_1 = C_1$.
	
	Note that 
	\bea
	\pi_r (\calP_n^c) 
	&\le& \pi_r \big( |S_{\delta_n} (\sg) | > s_n \big) + \pi_r ( \|\sg\|_{\max} > L_n )  \\ 
	&\le& \frac{\pi_r^u \big( |S_{\delta_n} (\sg) | > s_n \big) }{\pi_r^u \big(\sg \in \calU(\epsilon)  \big)  }  + \pi_r ( \lambda_{\max}(\sg) > L_n )  \\
	&\le& \frac{\pi_{r=0}^u \big( |S_{\delta_n} (\sg) | > s_n \big) }{\pi_r^u \big(\sg \in \calU(\epsilon)  \big)  }  + \pi_r ( \lambda_{\max}(\sg) > L_n )
	\eea
	where 
	$\pi_r ( \lambda_{\max}(\sg) > L_n ) \le \pi_r ( \lambda_{\max}(\sg) > \epsilon^{-1} )  =0 $ for all large $n$.
	Furthermore,  by Gershgorin circle theorem \cite{brualdi1994regions},
	\bea
	&& \pi_r^u \big(\sg \in \calU(\epsilon)  \big)  \\
	&\ge& \pi_r^u \Big( \min_j \big( \sigma_{jj} - \sum_{k\neq j} |\sigma_{kj}| \big) > 0 , \,\, \epsilon < \lambda_{\min}(\sg) \le \lambda_{\max}(\sg) <\epsilon^{-1}  \Big) \\
	&\ge& \pi_r^u \Big(  \epsilon < \min_j \big( \sigma_{jj} - \sum_{k\neq j} |\sigma_{kj}| \big)  \le 2 \max_j \sigma_{jj} < \epsilon^{-1} \Big)  \\
	&\ge& \pi_r^u \Big(  \epsilon < \min_j \big( \sigma_{jj} - \epsilon \big)  \le 2 \max_j \sigma_{jj} < \epsilon^{-1}  \,\,\Big|\,\, \max_{k\neq j} |\sigma_{kj}| < \frac{\epsilon}{p}  \Big) \pi_r^u \Big(  \max_{k\neq j} |\sigma_{kj}| <  \frac{\epsilon}{p}  \Big)  \\
	&\ge& \pi_{r=0}^u \Big(  \epsilon < \min_j \big( \sigma_{jj} - \epsilon \big)  \le 2 \max_j \sigma_{jj} < \epsilon^{-1}   \Big) \pi_{r=0}^u \Big(  \max_{k\neq j} |\sigma_{kj}| <  \frac{\epsilon}{p}  \Big)  \\
	&\ge& \exp \Big\{  - p \Big(  \frac{\lambda }{4 \epsilon }  + \log \frac{8\epsilon }{\lambda } \Big)  \Big\}  \exp \Big(  - \frac{2\sqrt{2} }{\sqrt{\pi^3} } \frac{\tau_1 p^3}{\epsilon}   \Big)   \\
	&\ge& \exp \Big\{  - p \Big(  \frac{\lambda }{4 \epsilon } + C + \log \frac{8\epsilon }{\lambda } \Big)  \Big\} \,\, \equiv \,\, \exp \big( - C' p \big)
	\eea
	for some positive constants $C$ and $C'$, where the fifth and sixth inequalities follow from Lemma \ref{lem:sigma_jk_ineq} and $\tau_1 = O\big( \frac{1}{p^2}\sqrt{\frac{s_0 \log p}{n}} \big)$, respectively.
	Because $\pi_{r=0}^u \big( |S_{\delta_n} (\sg) | > s_n \big)  \le \exp \big( - c_1 n \epsilon_n^2 / 3 \big)$ by Lemma \ref{lem:num_of_supp_bound}, we have
	\bea
	\pi_r (\calP_n^c ) &\le& \exp \Big\{ -  \frac{c_1 -1}{3} \, n\epsilon_n^2  \Big\} 
	\eea
	for all large $n$, 	which leads to \eqref{sieve_mass} by taking $(c_1-1)/3 = C_2 + 4$.
	Note that we can choose a constant $c_1$ as large as we wish.
	
	Finally, we show that \eqref{prior_thickness} holds and complete the proof.
	By Theorem 1, we have $\bbP_0 (S_0(\sg_0) \subseteq S_r(\hat{R}) ) \lra 1$ as $n\to\infty$.
	Let $A_n = \{ \bfX_n : S_0(\sg_0) \subseteq S_r(\hat{R}) \}$.
	Note that 
	\bea
	\bbE_0 \Big\{  \pi_r \big( \|\sg - \sg_0 \|_F^2  \ge M \epsilon_n^2 \big)   \Big\} 
	&\le& \bbP_0( A_n^c )  + \bbE_0 \Big\{  \pi_r \big( \|\sg - \sg_0 \|_F^2  \ge M \epsilon_n^2  \big) I_{A_n} \Big\}  ,
	\eea
	thus it suffices to show that \eqref{prior_thickness} holds on the event $A_n$.
	By Lemma 5.5 in \cite{lee2021betamixture}, we have
	\bean
	\pi_r \big( B_{\epsilon_n} \big) 
	&\ge& \pi_r \Big( \|\sg - \sg_0 \|_F^2 \le \frac{2\epsilon^4 \epsilon_0^2 }{3} \epsilon_n^2 \Big)  \nonumber \\
	&\ge& \pi_r \Big(  \max_{j \neq k} (\sigma_{jk}- \sigma_{0,jk})^2 \le \frac{2\epsilon^4 \epsilon_0^2 }{3}  \frac{s_0 \log p}{p(p-1) n}  ,\, \max_{1\le j \le p} (\sigma_{jj}- \sigma_{0,jj})^2 \le \frac{2\epsilon^4 \epsilon_0^2 }{3}  \frac{ \log p}{n}   \Big)   \nonumber \\
	&=:& \pi_r ( B_{n, \sg_0} ) .  \label{lb1}
	\eean
	By Weyl's theorem, for any $\sg \in B_{n, \sg_0}$ and $\sg_0 \in \calU(\epsilon_0)$, 
	\bea
	\lambda_{\min} (\sg) &\ge& \lambda_{\min}(\sg_0) - \|\sg- \sg_0\|_1 \\
	&\ge& \epsilon_0  - \sqrt{\frac{2\epsilon^4 \epsilon_0^2 }{3}  \frac{ \log p}{n} } -  p  \sqrt{\frac{2\epsilon^4 \epsilon_0^2 }{3}  \frac{s_0 \log p}{p(p-1) n}} \\
	&\ge& \epsilon , \\
	\lambda_{\max} (\sg)  &\le& \lambda_{\max} (\sg_0) +  \|\sg- \sg_0\|_1  \\
	&\le& \epsilon_0^{-1}  + \sqrt{\frac{2\epsilon^4 \epsilon_0^2 }{3}  \frac{ \log p}{n} } +  p  \sqrt{\frac{2\epsilon^4 \epsilon_0^2 }{3}  \frac{s_0 \log p}{p(p-1) n}}\\
	&\le& \epsilon^{-1}
	\eea
	for all large $n$, which implies $B_{n, \sg_0} \subset \calU(\epsilon)$ for any $\sg_0 \in \calU(\epsilon_0)$ and all large $n$.
	Thus, 
	\bean
	\pi_r \big( B_{\epsilon_n} \big)  &\ge& \pi_r^u \big( B_{\epsilon_n} \big)   \label{lb2}  \\
	&\ge& \pi_r^u \Big(  \max_{j \neq k} (\sigma_{jk}- \sigma_{0,jk})^2 \le \frac{2\epsilon^4 \epsilon_0^2 }{3}  \frac{s_0 \log p}{p(p-1) n}    \Big)  \nonumber \\
	&& \times \,\, \pi_r^u \Big(  \max_{1\le j \le p} (\sigma_{jj}- \sigma_{0,jj})^2 \le \frac{2\epsilon^4 \epsilon_0^2 }{3}  \frac{ \log p}{n}    \Big) . \nonumber
	\eean
	Note that
	\bea
	&& \pi_r^u \Big(  \max_{1\le j \le p} (\sigma_{jj}- \sigma_{0,jj})^2 \le \frac{2\epsilon^4 \epsilon_0^2 }{3}  \frac{ \log p}{n}    \Big)  \\
	&=& \prod_{j=1}^p \pi_r^u \Big(   (\sigma_{jj}- \sigma_{0,jj})^2 \le \frac{2\epsilon^4 \epsilon_0^2 }{3}  \frac{ \log p}{n}    \Big) \\
	&\ge& \prod_{j=1}^p  2 \sqrt{\frac{2\epsilon^4 \epsilon_0^2 }{3}  \frac{ \log p}{n} } \, \frac{\lambda}{2} \exp \Big\{  - \frac{\lambda}{2}\Big(   \sigma_{0,jj} +  \sqrt{\frac{2\epsilon^4 \epsilon_0^2 }{3}  \frac{ \log p}{n} }   \Big)  \Big\}  \\
	&\ge&  \bigg[ \lambda \sqrt{\frac{2\epsilon^4 \epsilon_0^2 }{3}  \frac{ \log p}{n} }   \exp \Big\{  - \frac{\lambda}{2}\Big(   \epsilon_0^{-1} +  \sqrt{\frac{2\epsilon^4 \epsilon_0^2 }{3}  \frac{ \log p}{n} }   \Big)  \Big\}  \bigg]^p  \\ 
	&\ge&  \exp \Big\{  - p \lambda \epsilon_0^{-1}  - p \log \Big( \frac{ \sqrt{3n}  }{ \lambda \sqrt{2\epsilon^4 \epsilon_0^2 \log p}   }  \Big)     \Big\}  \\
	&\ge& \exp \Big[   - p \log p -  \{ 1 +(2\beta)^{-1}  \} p \log p     \Big]    \\
	&\ge& \exp \Big[  - \big\{  2 + (2\beta)^{-1}) \big\} p \log p  \Big] 
	\eea
	for all large $n$.
	Furthermore, on the event $A_n$,
	\bean
	&& \pi_r^u \Big(  \max_{j \neq k} (\sigma_{jk}- \sigma_{0,jk})^2 \le \frac{2\epsilon^4 \epsilon_0^2 }{3}  \frac{s_0 \log p}{p(p-1) n}    \Big)  \nonumber \\
	&=& \prod_{(j,k) \in \hat{S}_r } \pi_r^u \Big(   (\sigma_{jk}- \sigma_{0,jk})^2 \le \frac{2\epsilon^4 \epsilon_0^2 }{3}  \frac{s_0 \log p}{p(p-1) n}    \Big) \nonumber \\
	&\ge& \prod_{(j,k) \in S_0(\sg_0) } \pi_r^u \Big(   (\sigma_{jk}- \sigma_{0,jk})^2 \le \frac{2\epsilon^4 \epsilon_0^2 }{3}  \frac{s_0 \log p}{p(p-1) n}    \Big)   \label{sig_S0} \\
	&& \times \,\, \prod_{(j,k) \in S_0(\sg_0)^c \cap \hat{S}_r  } \pi_r^u \Big(   \sigma_{jk}^2 \le \frac{2\epsilon^4 \epsilon_0^2 }{3}  \frac{s_0 \log p}{p(p-1) n}    \Big)  . \label{sig_not_S0}
	\eean
	Note that, for ${\pi}^u(\sigma_{jk}) = \int_0^1 {\pi}^u(\sigma_{jk} \mid \psi_{jk} )\pi^u(\psi_{jk}) d\psi_{jk} $,
	\bea
	&& \prod_{(j,k) \in S_0(\sg_0) }  {\pi}_r^u \Big(   (\sigma_{jk}- \sigma_{0,jk})^2 \le \frac{2\epsilon^4 \epsilon_0^2 s_0 \log p}{3 p(p-1) n}    \Big)   \\
	&\ge&  \prod_{(j,k) \in S_0(\sg_0) }  {\pi}^u \bigg( \sigma_{0,jk} + \sqrt{\frac{2\epsilon^4 \epsilon_0^2 s_0 \log p}{3 p(p-1) n}  }  \, \bigg) \, 2 \sqrt{\frac{2\epsilon^4 \epsilon_0^2 s_0 \log p }{3 p(p-1) n}  }   \\
	&\ge& \bigg\{   2  \pi^u ( 2\epsilon_0^{-2} )  \sqrt{\frac{2\epsilon^4 \epsilon_0^2 s_0 \log p }{3 p(p-1) n}   }    \bigg\}^{s_0 }   \\
	&\ge& \exp \bigg[  s_0 \log \big\{ \pi^u (2 \epsilon_0^{-2})  \big\}   - \frac{1}{2} s_0 \log \Big(  \frac{3 p^2 n}{2\epsilon^4 \epsilon_0^2 s_0 \log p }  \Big)  \bigg]  \\
	&\ge& \exp \bigg\{   s_0 \log \Big( \frac{\tau_1 \epsilon_0^4 }{4 \sqrt{2 \pi^3 } } \Big)  - \frac{1}{2} s_0 \log (  p^2 n ) + \frac{1}{2} s_0 \log \Big( \frac{2\epsilon^4 \epsilon_0^2 s_0 \log p}{3} \Big)  \bigg\}  \\
	&\ge& \exp \bigg\{  -  s_0 \log (p^2 \sqrt{n})  -   s_0 \log (p \sqrt{n} )\bigg\} \\
	&\ge& \exp \Big\{  -  \big(   3 + \beta^{-1}  \big) s_0 \log p  \Big\} ,
	\eea
	where the fourth and fifth inequalities follow from Lemma 5.6 in \cite{lee2021betamixture} and $\tau_1 \gtrsim (p^2 \sqrt{n})^{-1}$, respectively.
	Also note that 
	\bea
	&&  \prod_{(j,k) \in S_0(\sg_0)^c \cap \hat{S}_r }  {\pi}_r^u \Big(  \sigma_{jk}^2 \le \frac{2\epsilon^4 \epsilon_0^2 s_0 \log p }{3 p(p-1) n}    \Big)   \\
	&=& \prod_{(j,k) \in S_0(\sg_0)^c \cap \hat{S}_r }   \bigg\{   1-  {\pi}_r^u \Big(  \sigma_{jk}^2 > \frac{2\epsilon^4 \epsilon_0^2 s_0 \log p }{3 p(p-1) n}    \Big)   \bigg\}  \\
	&\ge&  \bigg\{     1-  \frac{2\sqrt{2}}{\sqrt{\pi^3} } \tau_1  \sqrt{\frac{3 p(p-1) n}{2\epsilon^4 \epsilon_0^2 s_0 \log p } } \bigg\}^{p(p-1)/2 }   \\
	&\ge&  \exp  \bigg\{   -  2\tau_1 p^2  \sqrt{\frac{3p^2 n }{\pi^3 \epsilon^4 \epsilon_0^2 s_0 \log p } }  \bigg\}  \\
	&\ge&  \exp  \big(   -  C p  \big) 
	\eea
	for some constant $C>0$, 
	by inequality \eqref{lb3} in the proof of Lemma \ref{lem:sigma_jk_ineq} and the condition $\tau_1 \lesssim p^{-2} \sqrt{(s_0 \log p) /n}$.
	Thus,  for some constant $C>0$,
	\bea
	\pi_r ( B_{\epsilon_n} )  &\ge& \exp \Big[   - \big\{  2+ (2\beta)^{-1} \big\} p \log p - \big(  3+ \beta^{-1} \big) s_0 \log p - C p   \Big] \\
	&\ge& \exp \Big\{  - \big(  3+ \beta^{-1} \big) n \epsilon_n^2  \Big\}  ,
	\eea
	which leads to \eqref{prior_thickness} by taking $3+ \beta^{-1} = C_2$.  \hfill $\blacksquare$

\section*{Appendix B. Concentration inequalities for correlation coefficients}\label{app:concentration_corr}

Let $X_i = (X_{i1} , X_{i2})^T \in \bbR^2, i=1,\ldots, n$ be random samples from the bivariate normal distribution with zero mean vector and $\V(X_{i1} ) = \sigma_1^2$, $\V(X_{i2} ) = \sigma_2^2$ and $\C(X_{i1}, X_{i2}) = \rho \sigma_1 \sigma_2$.
Let 
\bea
\hat{\rho} &=& \frac{\sum_{i=1}^n X_{i1} X_{i2} }{\sqrt{\sum_{i=1}^n X_{i1}^2 \sum_{i=1}^n X_{i2}^2} }
\eea
be the sample correlation coefficient.

\noindent
\begin{lemma}\label{lem:samp_corr_concen}
If $\rho=0$, then for any $r>0$, we have 
\bea
\bbP_{\rho=0} (|\hat{\rho}| \ge r ) &\le& \frac{4}{r\sqrt{n} }\exp \Big( - \frac{n}{2}r^2\Big).
\eea
\end{lemma}

\noindent
{\bf Proof}. By Theorem 4.2.1 of \cite{anderson2003introduction} (page 120), the density function of $\hat{\rho}$ is given by 
	\bea
	f( x \mid \rho=0  ) &=& \frac{\Gamma\big( \frac{n}{2} \big) }{\Gamma\big( \frac{n-1}{2} \big) \sqrt{\pi} } (1- x^2)^{ \frac{n-3}{2} }  , \quad -1\le x \le 1.
	\eea
	Because the distribution of $\hat{\rho}$ is symmetric about 0, we have
	\bea
	\bbP_0 ( | \hat{\rho}| \ge r)  &=& 2 \bbP_0 (\hat{\rho}  \ge r ) .
	\eea
	Note that 
	\bea
	\bbP_0 (\hat{\rho} \ge r )  &\le&  \int_r^1  \frac{\Gamma\big( \frac{n}{2} \big) }{\Gamma\big( \frac{n-1}{2} \big) \sqrt{\pi} }  (1- x^2)^{\frac{n-3}{2} } dx \\
	&\le& \frac{\Gamma\big( \frac{n}{2} \big) }{\Gamma\big( \frac{n-1}{2} \big) \sqrt{\pi} }  \int_r^1 \exp \Big(  - \frac{n}{2}x^2 + \frac{3}{2} x^2   \Big) dx ,
	\eea
	where the second inequality because $\log(1-x) \le -x$ for all $x \in \bbR$.
	By Watson’s monotonicity for gamma functions \citep{watson1959note}, the last display is bounded above by
	\bea
	&& \sqrt{\frac{n-1}{2 \pi }}  \int_r^1 \exp \Big(  - \frac{n}{2}x^2 + \frac{3}{2} x^2   \Big) dx \\
	&\le& \sqrt{\frac{n-1}{2 \pi }}  e^{\frac{3}{2}}  \int_r^1 \exp \Big(  - \frac{n}{2}x^2 \Big) dx  \\
	&=& \sqrt{\frac{n-1}{n }}  e^{\frac{3}{2}}    \int_r^1 \Big( \frac{2\pi}{n} \Big)^{- \frac{1}{2} } \exp \Big( - \frac{n}{2} x^2 \Big) dx  \\
	&\le& e^{\frac{3}{2}}  \int_{r\sqrt{n} }^\infty (2\pi)^{-\frac{1}{2} } \exp \Big( - \frac{1}{2}z^2 \Big)  dz   \\
	&\le& \frac{e^{\frac{3}{2}}   }{\sqrt{2\pi} }  \frac{1}{r \sqrt{n}}   \exp \Big( - \frac{n}{2} r^2 \Big)  \\
	&\le& \frac{2}{r\sqrt{n} }  \exp \Big( - \frac{n}{2} r^2 \Big)  ,
	\eea
	where the last inequality holds because $e^{3/2}/\sqrt{2\pi} < 2$.
	This gives the desired result. \hfill$\blacksquare$

\begin{lemma}\label{lem:samp_corr_small_prob}
    For any $1\ge |\rho| > r > 0$, $r < 1/2$ and $n>2$, 
	\bean\label{ineq:rho}
	\bbP_\rho (|\hat{\rho}|  \le r)  &\le&   \frac{23 \, ( n |\rho| r \wedge 1)}{\sqrt{n} ( |\rho| - r) }  \exp \Big\{  - \frac{n (|\rho|-r)^2  }{2 (1+ \rho^2 )}    \Big\}   .
	\eean
\end{lemma}

\noindent
{\bf Proof}. Note that by Theorem 4.2.2 of \cite{anderson2003introduction} (pages 125--126), the density of $\hat{\rho}$ is given by 
	\bea
	f( x \mid  \rho )  &=& \frac{n-1}{\sqrt{2\pi}} \frac{\Gamma(n) }{\Gamma\big( n + \frac{1}{2} \big)} (1- \rho^2)^{\frac{n}{2}}  (1-x^2 )^{\frac{n-3}{2} }  ( 1- \rho x)^{-n + \frac{1}{2} } {}_{2}F_1 \Big(  \frac{1}{2} , \frac{1}{2} ; n +\frac{1}{2} ; \frac{1+\rho x}{2} \Big) ,
	\eea
	where $-1\le x\le 1$ and ${}_{2}F_1(\cdot,\cdot; \cdot;\cdot)$ is a hypergeometric function.
	Thus, it suffices to show that \eqref{ineq:rho} holds when $\rho > r >0$, because we have $f(x \mid -\rho)  = f(-x \mid \rho)$.
	
	Note that for any $m\ge 0$,
	\bea
	{}_{2}F_1 \Big(  \frac{1}{2} , \frac{1}{2} ; n +\frac{1}{2} ; \frac{1+\rho x}{2} \Big)  
	&=&  \sum_{k=0}^\infty \frac{ \big\{ \big(\frac{1}{2}\big)_k \big\}^2 }{ \big( n +\frac{1}{2}\big)_k   \, k!} \Big( \frac{1+\rho x}{2} \Big)^k   \\
	&\equiv&  \sum_{k=0}^\infty b_k  \Big( \frac{1+\rho x}{2} \Big)^k  \\
	&=:&  S_m(x)  + R_m(x) ,
	\eea
	where $S_m(x)  = \sum_{k=0}^m b_k  \Big( \frac{1+\rho x}{2} \Big)^k$ and $R_m(x)  = \sum_{k=m+1}^\infty b_k  \Big( \frac{1+\rho x}{2} \Big)^k$.	
	Then $b_0=1$ and $b_1 = 1/\{2(n+1)\}$.
	Let $C_n := (n-1) \Gamma(n) /\{ \sqrt{2\pi} \Gamma(n +1/2) \}$ and $g_0(x) := (1- \rho^2)^{\frac{n}{2}}  (1-x^2 )^{\frac{n-3}{2} }  ( 1- \rho x)^{-n + \frac{1}{2} }  $. 
	Then,
	\bea
	&& \bbP_\rho (-r \le \hat{\rho} \le r)  \\
	&=& C_n \int_{-r}^r  g_0(x) \big\{ S_m(x)  + R_m(x)  \big\}  dx  \\
	&\le& C_n \int_{-r}^r g_0(x) S_m(x) dx +  C_n \int_{-r}^r g_0(x) b_m \Big( \frac{1+\rho x}{2} \Big)^m \frac{1+\rho x}{1- \rho x} dx  \\
	&\le& C_n \int_{-r}^r g_0(x) S_m(x) dx +  C_n \int_{-r}^r g_0(x) b_m \Big( \frac{1+\rho x}{2} \Big)^m \frac{1+\rho r}{1- \rho r} dx ,
	\eea
	where the first inequality holds because $b_k \le b_m$ for any $k >m$.
	By taking $m=1$, the last display becomes 
	\bea
	&& C_n \Big[ \int_{-r}^r g_0(x) \Big\{ 1 + \frac{1+\rho x}{4(n+1)}  \Big\} dx   + \int_{-r}^r g_0(x) \frac{1+\rho x}{4(n+1)}  \frac{1+\rho r}{1- \rho r} dx  \Big]  \\
	&\le&  \Big\{  1+  \frac{1+ \rho r}{2(n+1)(1-\rho r)}   \Big\}   \, C_n  \int_{-r}^r g_0(x) dx  \\
	&\le&  \Big\{  1+  \frac{1+ \rho r}{2(n+1)(1-\rho r)}   \Big\}  \sqrt{\frac{n}{2\pi} }  \int_{-r}^r g_0(x) dx ,
	\eea
	because $C_n \le \sqrt{(n-1)/(2\pi)}$ \citep{watson1959note}.
	
	Let $M_0 := \int_{-r}^r g_0(x) dx$.  
	Then,
	\bea
	\sqrt{n} M_0 
	&=& \sqrt{n} \int_{-r}^r \Big( \frac{1-x^2}{1- \rho x} \Big)^{\frac{n-3}{2} } \Big( \frac{1-\rho^2}{1-\rho x} \Big)^{\frac{n}{2} } \frac{1}{1- \rho x} dx  \\
	&\le& \frac{\sqrt{n}}{1-\rho r} \int_{-r}^r  \Big( \frac{1-x^2}{1- \rho x} \Big)^{\frac{n-3}{2} } \Big( \frac{1-\rho^2}{1-\rho x} \Big)^{\frac{n}{2} }  dx \\
	&=& \frac{\sqrt{n}}{1-\rho r} \int_{-r}^r   \Big( 1 - \frac{x^2 - \rho x}{1- \rho x} \Big)^{\frac{n-3}{2} }   \Big( 1 + \frac{\rho x-\rho^2}{1-\rho x} \Big)^{\frac{n}{2} }  dx \\
	&\le& \frac{\sqrt{n}}{1-\rho r} \int_{-r}^r     \exp \Big\{   - \frac{n (x-\rho)^2}{2(1-\rho x)}   + \frac{3 x(x-\rho)}{2(1-\rho x)}  \Big\}  dx  \\
	&\le& \frac{\sqrt{n}}{1-\rho r}  \exp \Big\{  \frac{3r(r+\rho)}{2(1-\rho r) }   \Big\}  \int_{-r}^r     \exp \Big\{  -\frac{n (x-\rho)^2}{2(1+\rho r)}  \Big\} dx  \\
	&=& \frac{\sqrt{2 \pi (1+\rho r)}}{1-\rho r}  \exp \Big\{  \frac{3r(r+\rho)}{2(1-\rho r) }   \Big\}  \, \bbP_0 \bigg( \frac{\sqrt{n}(\rho -r) }{\sqrt{1+\rho r} }  \le Z  \le  \frac{\sqrt{n}(\rho + r) }{\sqrt{1+\rho r} }  \bigg) ,
	\eea
	where the second inequality holds because $\log(1-x) \le -x$ for all $x \in \bbR$.
	Let $a_{n1} := \sqrt{n}(\rho -r)/\sqrt{1+\rho r}$, $a_{n2} := \sqrt{n}(\rho +r)/\sqrt{1+\rho r}$ and 
	\bea
	\tilde{C} &:=&  \frac{\sqrt{2 \pi (1+\rho r)}}{1-\rho r}  \exp \Big\{  \frac{3r(r+\rho)}{2(1-\rho r) }   \Big\}  .
	\eea
	Then,
	\bea
	\sqrt{n} M_0 
	&\le& \tilde{C}  \int_{a_{n1}}^{a_{n2}} \frac{1}{\sqrt{2\pi} } e^{- \frac{t^2}{2}} dt  \\
	&\le& \tilde{C}  \frac{1}{\sqrt{2\pi} a_{n1}} \int_{a_{n1}}^{a_{n2}}  t e^{- \frac{t^2}{2}} dt  \\
	&=& \frac{\tilde{C}  }{\sqrt{2\pi} a_{n1}}  \Big( e^{-\frac{a_{n1}^2}{2} }  - e^{-\frac{a_{n2}^2}{2} } \Big)  \\
	&=& \frac{1+\rho r}{\sqrt{n}(1-\rho r)(\rho -r)}   \exp \Big\{  \frac{3r(r+\rho)}{2(1-\rho r) }  - \frac{n(\rho-r)^2}{2(1+\rho r)} \Big\} \Big\{ 1 - \exp \Big( - \frac{2n \rho r}{1+\rho r} \Big)   \Big\}  \\
	&\le& \frac{1+\rho r}{\sqrt{n}(1-\rho r)(\rho -r)}   \exp \Big\{  \frac{3r(r+\rho)}{2(1-\rho r) }  - \frac{n(\rho-r)^2}{2(1+\rho r)} \Big\} \Big(  \frac{2n \rho r}{1+\rho r} \wedge 1 \Big) .
	\eea
	Therefore, we have 
	\bea
	\bbP_\rho ( -r \le \hat{\rho} \le r)  
	&\le&  \Big\{  1+  \frac{1+ \rho r}{2(n+1)(1-\rho r)}   \Big\}  \sqrt{\frac{n}{2\pi} }  M_0 \\
	&\le& \frac{3e^{\frac{9}{4} }    }{\sqrt{2\pi} }     \frac{( 2 n\rho r \wedge 2 )}{\sqrt{n}(\rho  -r) }  \exp \Big\{  - \frac{n(\rho-r)^2}{2(1+\rho r)} \Big\}  \\
	&\le& \frac{23 \, ( n \rho r \wedge 1)}{\sqrt{n} ( \rho - r) }  \exp \Big\{  - \frac{n (\rho-r)^2  }{2 (1+ \rho^2 )}    \Big\} 
	\eea
	because $0<r<1/2$ and $n>2$.  \hfill$\blacksquare$

\section*{Appendix C. Auxiliary results}\label{app:auxil}
\noindent
\begin{lemma}\label{lem:sigma_jk_ineq}
    For a given constant $0< \epsilon < 1/3$, we have 
	\bea
	\pi_{r=0}^u \Big( 2\epsilon < \sigma_{jj} <   (2\epsilon)^{-1}  \text{ for all } j=1,\ldots, p \Big) 
	&\ge& \exp \Big\{  - p \Big(  \frac{\lambda }{4 \epsilon } + \log \frac{8\epsilon }{\lambda } \Big)  \Big\} \\
	\pi_{r=0}^u \Big(  \max_{k\neq j} |\sigma_{kj}| <  \frac{\epsilon}{p}  \Big)  
	&\ge& \exp \Big(  - \frac{2\sqrt{2} }{\sqrt{\pi^3} } \frac{\tau_1 p^3}{\epsilon}   \Big)  .
	\eea
\end{lemma}
\noindent
{\bf Proof}. Because $0<\epsilon <1/3$ and the probability density of an exponential distribution is a decreasing function,
	\bea
	&&\pi_{r=0}^u \Big( 2\epsilon < \sigma_{jj} <   (2\epsilon)^{-1}  \text{ for all } j=1,\ldots, p \Big)   \\
	&=& \prod_{j=1}^p \pi_{r=0}^u \Big( 2\epsilon < \sigma_{jj} <   (2\epsilon)^{-1}   \Big)   \\
	&\ge& \Big[  \big\{  (2\epsilon)^{-1}  - 2\epsilon \big\} \frac{\lambda}{2} \exp \Big( - \frac{\lambda }{4 \epsilon} \Big)  \Big]^p  \\
	&\ge& \exp \Big\{  - p \Big(  \frac{\lambda }{4 \epsilon } + \log \frac{8\epsilon }{\lambda } \Big)  \Big\} .
	\eea
	
	Note that 
	\bean
	\pi_{r=0}^u \big( |\sigma_{kj}| \ge \epsilon / p \big) 
	&\le& \frac{1}{\tau_1} \sqrt{\frac{2}{\pi^3} } \int_{\epsilon / p}^{\infty} \log \Big( 1 + \frac{2 \tau_1^2}{x^2} \Big)  dx  \nonumber \\
	&\le& \sqrt{\frac{2}{\pi^3} }  \int_{\epsilon / p}^{\infty} \frac{2 \tau_1}{x^2} dx  \nonumber \\
	&=&  \frac{2\sqrt{2} }{\sqrt{\pi^3} } \frac{\tau_1 p}{\epsilon}  ,  \label{lb3}
	\eean
	where the first inequality follows from Theorem 1 in \cite{carvalho2010horseshoe}.
	Therefore,
	\bea
	\pi_{r=0}^u \Big(  \max_{k\neq j} |\sigma_{kj}| <  \frac{\epsilon}{p}  \Big)  
	&=& \prod_{k\neq j} \Big\{  1 - \pi_{r=0}^u \big( |\sigma_{kj}| \ge \epsilon / p \big)   \Big\} \\
	&\ge& \Big(  1 - \frac{2\sqrt{2} }{\sqrt{\pi^3} } \frac{\tau_1 p}{\epsilon}  \Big)^{p(p-1)/2} \\
	&\ge& \exp \Big(  - \frac{2\sqrt{2} }{\sqrt{\pi^3} } \frac{\tau_1 p^3}{\epsilon}   \Big) ,
	\eea
 which completes the proof. \hfill $\blacksquare$

\begin{lemma}\label{lem:num_of_supp_bound}
    If $\delta_n = \epsilon^3 \epsilon_n$, 
	$s_n \le c_1 n \epsilon_n^2 / \log p$,
	$p^2 \tau_1 < s_n \delta_n$, 
	$\tau_1 = o(\delta_n)$ and 
	$s_n \log \{2s_n\delta_n / (p^2 \tau_1) \} \ge c_1 n \epsilon_n^2/2$, we have
	\bea
	\pi_{r=0}^u \big( |S_{\delta_n} (\sg) | > s_n \big)  &\le& \exp \Big(   - \frac{c_1 n \epsilon_n^2}{3}   \Big) 
	\eea
	for all large $n$.
\end{lemma}

\noindent
{\bf Proof}. For any $1\le k\neq j \le p$, 
	\bea
	\nu_n &\equiv& \pi_{r=0}^u (|\sigma_{kj}| > \delta_n ) \\
	&\le& \frac{2\sqrt{2}}{\sqrt{\pi^3}} \frac{\tau_1}{\delta_n}  \\
	&\le& \frac{\tau_1}{\delta_n}  , 
	\eea
	where the first inequality follows from Theorem 1 in \cite{carvalho2010horseshoe}.
	Note that $\binom{p}{2}  \nu_n \le p(p-1) \tau_1/ (2\delta_n) < s_n$.
	Then, by Lemma A.3 in \cite{song2018nearly}, 
	\bea
	\pi_{r=0}^u \big( |S_{\delta_n} (\sg) | > s_n \big)
	&\le& 1- \Phi \left[   \sqrt{ 2 \binom{p}{2}   H \Big\{   \nu_n , \,  s_n / \binom{p}{2}  \Big\}  }  \right]   ,
	\eea
	where $\Phi$ is the cdf of $N(0,1)$ and 
	$H(a , b) := b \log (b/a) + (1-b) \log\{(1-b)/(1-a) \}$.
	Furthermore,
	\bea
	1- \Phi \left[   \sqrt{ 2 \binom{p}{2}   H \Big\{   \nu_n , \,  s_n / \binom{p}{2}  \Big\}  }  \right] 
	&\le& \frac{\exp \Big[ - \binom{p}{2}   H \Big\{   \nu_n , \,  s_n / \binom{p}{2}  \Big\}   \Big] }{\sqrt{2\pi} \sqrt{2  \binom{p}{2}   H \Big\{   \nu_n , \,  s_n / \binom{p}{2}  \Big\} }  }
	\eea
	and 
	\bea
	\binom{p}{2}   H \Big\{   \nu_n , \,  s_n / \binom{p}{2}  \Big\}
	&=& s_n \log \Big\{  \frac{s_n}{\binom{p}{2} \nu_n }  \Big\}
	+  \Big\{ \binom{p}{2}-  s_n \Big\} \log \Big\{  \frac{\binom{p}{2}-   s_n}{\binom{p}{2}- \binom{p}{2} \nu_n }  \Big\}  .
	\eea
	
	Note that
	\bea
	s_n \log \Big\{  \frac{s_n}{\binom{p}{2} \nu_n }  \Big\}
	&\ge& s_n \log \Big(  \frac{2s_n\delta_n}{p^2 \tau_1} \Big)  \,\,\ge\,\, c_1 n \epsilon_n^2/2
	\eea
	and
	\bea
	\Big\{ \binom{p}{2}-  s_n \Big\} \log \Big\{  \frac{\binom{p}{2}-   s_n}{\binom{p}{2}- \binom{p}{2} \nu_n }  \Big\}
	&\ge& - 2\Big\{  \binom{p}{2}-   s_n \Big\} \frac{s_n - \binom{p}{2}\nu_n }{\binom{p}{2} (1-\nu_n)}   \\
	&\ge& - \frac{2s_n}{1-\nu_n } \\
	&\ge& - 3 s_n  \\
	&\ge& -  \frac{3 c_1 n \epsilon_n^2}{\log p}
	\eea
	for all large $n$ because $\tau_1 = o(\delta_n)$.
	
	Therefore,
	\bea
	\pi_{r=0}^u \big( |S_{\delta_n} (\sg) | > s_n \big)
	&\le& \exp \Big(   - \frac{c_1 n \epsilon_n^2}{2} + \frac{3 c_1 n \epsilon_n^2}{\log p}  \Big) \\
	&\le& \exp \Big(   - \frac{c_1 n \epsilon_n^2}{3}   \Big) 
	\eea
	for all large $n$. \hfill $\blacksquare$

\vskip 0.2in
\bibliography{twostepCOV}

\begin{thebibliography}{33}
\providecommand{\natexlab}[1]{#1}
\providecommand{\url}[1]{\texttt{#1}}
\expandafter\ifx\csname urlstyle\endcsname\relax
  \providecommand{\doi}[1]{doi: #1}\else
  \providecommand{\doi}{doi: \begingroup \urlstyle{rm}\Url}\fi

\bibitem[son()]{song2017nearly}


\bibitem[Alon et~al.(1999)Alon, Barkai, Notterman, Gish, Ybarra, Mack, and
  Levine]{alon1999}
U.~Alon, N.~Barkai, D.~A. Notterman, K.~Gish, S.~Ybarra, D.~Mack, and A.~J.
  Levine.
\newblock Broad patterns of gene expression revealed by clustering analysis of
  tumor and normal colon tissues probed by oligonucleotide arrays.
\newblock \emph{Proc. Natl. Acad. Sci.}, 96\penalty0 (12):\penalty0 6745--6750,
  1999.

\bibitem[Anderson(2003)]{anderson2003introduction}
T.W. Anderson.
\newblock \emph{An Introduction to Multivariate Statistical Analysis}.
\newblock Wiley Series in Probability and Statistics. Wiley, 2003.
\newblock ISBN 9780471360919.

\bibitem[Bickel and Levina(2008)]{bickel2008covariance}
Peter~J Bickel and Elizaveta Levina.
\newblock Covariance regularization by thresholding.
\newblock \emph{The Annals of Statistics}, 36\penalty0 (6):\penalty0
  2577--2604, 2008.

\bibitem[Bien and Tibshirani(2011)]{bien2011sparse}
Jacob Bien and Robert~J Tibshirani.
\newblock Sparse estimation of a covariance matrix.
\newblock \emph{Biometrika}, 98\penalty0 (4):\penalty0 807--820, 2011.

\bibitem[Brualdi and Mellendorf(1994)]{brualdi1994regions}
Richard~A Brualdi and Stephen Mellendorf.
\newblock Regions in the complex plane containing the eigenvalues of a matrix.
\newblock \emph{The American mathematical monthly}, 101\penalty0 (10):\penalty0
  975--985, 1994.

\bibitem[Cai and Liu(2011)]{cai2011adaptive}
Tony Cai and Weidong Liu.
\newblock Adaptive thresholding for sparse covariance matrix estimation.
\newblock \emph{Journal of the American Statistical Association}, 106\penalty0
  (494):\penalty0 672--684, 2011.

\bibitem[Carvalho et~al.(2010)Carvalho, Polson, and
  Scott]{carvalho2010horseshoe}
Carlos~M Carvalho, Nicholas~G Polson, and James~G Scott.
\newblock The horseshoe estimator for sparse signals.
\newblock \emph{Biometrika}, 97\penalty0 (2):\penalty0 465--480, 2010.

\bibitem[Castillo et~al.(2015)Castillo, Schmidt-Hieber, Van~der Vaart,
  et~al.]{castillo2015bayesian}
Isma{\"e}l Castillo, Johannes Schmidt-Hieber, Aad Van~der Vaart, et~al.
\newblock Bayesian linear regression with sparse priors.
\newblock \emph{Annals of Statistics}, 43\penalty0 (5):\penalty0 1986--2018,
  2015.

\bibitem[El~Karoui(2008)]{el2008operator}
Noureddine El~Karoui.
\newblock Operator norm consistent estimation of large-dimensional sparse
  covariance matrices.
\newblock \emph{The Annals of Statistics}, 36\penalty0 (6):\penalty0
  2717--2756, 2008.

\bibitem[Fan and Lv(2008)]{fan2008sure}
Jianqing Fan and Jinchi Lv.
\newblock Sure independence screening for ultrahigh dimensional feature space.
\newblock \emph{Journal of the Royal Statistical Society: Series B (Statistical
  Methodology)}, 70\penalty0 (5):\penalty0 849--911, 2008.

\bibitem[Ghosal et~al.(2000)Ghosal, Ghosh, and Van
  Der~Vaart]{ghosal2000convergence}
Subhashis Ghosal, Jayanta~K Ghosh, and Aad~W Van Der~Vaart.
\newblock Convergence rates of posterior distributions.
\newblock \emph{Annals of Statistics}, 28\penalty0 (2):\penalty0 500--531,
  2000.

\bibitem[Jeffreys(1998)]{jeffreys1998theory}
Harold Jeffreys.
\newblock \emph{The theory of probability}.
\newblock OUP Oxford, 1998.

\bibitem[Khan et~al.(2001)Khan, Wei, Ringner, Saal, Ladanyi, Westermann,
  Berthold, Schwab, Antonescu, Peterson, et~al.]{khan2001classification}
Javed Khan, Jun~S Wei, Markus Ringner, Lao~H Saal, Marc Ladanyi, Frank
  Westermann, Frank Berthold, Manfred Schwab, Cristina~R Antonescu, Carsten
  Peterson, et~al.
\newblock Classification and diagnostic prediction of cancers using gene
  expression profiling and artificial neural networks.
\newblock \emph{Nature medicine}, 7\penalty0 (6):\penalty0 673--679, 2001.

\bibitem[Khare and Rajaratnam(2011)]{khare2011wishart}
Kshitij Khare and Bala Rajaratnam.
\newblock Wishart distributions for decomposable covariance graph models.
\newblock \emph{The Annals of Statistics}, 39\penalty0 (1):\penalty0 514--555,
  2011.

\bibitem[Lam and Fan(2009)]{lam2009sparsistency}
Clifford Lam and Jianqing Fan.
\newblock Sparsistency and rates of convergence in large covariance matrix
  estimation.
\newblock \emph{Annals of statistics}, 37\penalty0 (6B):\penalty0 4254, 2009.

\bibitem[Lee and Cao(2020)]{lee2020bayesian}
Kyoungjae Lee and Xuan Cao.
\newblock Bayesian group selection in logistic regression with application to
  {MRI} data analysis.
\newblock \emph{Biometrics}, 2020.

\bibitem[Lee et~al.(2021)Lee, Jo, and Lee]{lee2021betamixture}
Kyoungjae Lee, Seongil Jo, and Jaeyong Lee.
\newblock The beta-mixture shrinkage prior for sparse covariances with
  posterior minimax rates, 2021.

\bibitem[Liu and Martin(2019)]{liu2019empirical}
Chang Liu and Ryan Martin.
\newblock An empirical $ g $-{Wishart} prior for sparse high-dimensional
  gaussian graphical models.
\newblock \emph{arXiv preprint arXiv:1912.03807}, 2019.

\bibitem[Ly et~al.(2016)Ly, Verhagen, and Wagenmakers]{ly2016harold}
Alexander Ly, Josine Verhagen, and Eric-Jan Wagenmakers.
\newblock Harold {Jeffreys}’s default {Bayes} factor hypothesis tests:
  Explanation, extension, and application in psychology.
\newblock \emph{Journal of Mathematical Psychology}, 72:\penalty0 19--32, 2016.

\bibitem[Martin et~al.(2017)Martin, Mess, and Walker]{martin2017empirical}
Ryan Martin, Raymond Mess, and Stephen~G Walker.
\newblock Empirical {Bayes} posterior concentration in sparse high-dimensional
  linear models.
\newblock \emph{Bernoulli}, 23\penalty0 (3):\penalty0 1822--1847, 2017.

\bibitem[Maurya(2016)]{maurya2016well}
Ashwini Maurya.
\newblock A well-conditioned and sparse estimation of covariance and inverse
  covariance matrices using a joint penalty.
\newblock \emph{The Journal of Machine Learning Research}, 17\penalty0
  (1):\penalty0 4457--4484, 2016.

\bibitem[Polson and Scott(2010)]{polson2010shrink}
Nicholas~G Polson and James~G Scott.
\newblock Shrink globally, act locally: Sparse {Bayesian} regularization and
  prediction.
\newblock \emph{Bayesian statistics}, 9\penalty0 (501-538):\penalty0 105, 2010.

\bibitem[Rothman et~al.(2008)Rothman, Bickel, Levina, and Zhu]{rothman08}
A.~J. Rothman, P.~J. Bickel, E.~Levina, and J.~Zhu.
\newblock Sparse permutation invariant covariance estimation.
\newblock \emph{Electron. J. Stat.}, 2:\penalty0 494--515, 2008.

\bibitem[Rothman(2012)]{rothman2012positive}
Adam~J Rothman.
\newblock Positive definite estimators of large covariance matrices.
\newblock \emph{Biometrika}, 99\penalty0 (3):\penalty0 733--740, 2012.

\bibitem[Rothman et~al.(2009)Rothman, Levina, and Zhu]{rothman2009generalized}
Adam~J Rothman, Elizaveta Levina, and Ji~Zhu.
\newblock Generalized thresholding of large covariance matrices.
\newblock \emph{Journal of the American Statistical Association}, 104\penalty0
  (485):\penalty0 177--186, 2009.

\bibitem[Silva and Ghahramani(2009)]{silva2009hidden}
Ricardo Silva and Zoubin Ghahramani.
\newblock The hidden life of latent variables: {Bayesian} learning with mixed
  graph models.
\newblock \emph{Journal of Machine Learning Research}, 10\penalty0
  (Jun):\penalty0 1187--1238, 2009.

\bibitem[Song and Liang(2015)]{song2015split}
Qifan Song and Faming Liang.
\newblock A split-and-merge {Bayesian} variable selection approach for
  ultrahigh dimensional regression.
\newblock \emph{Journal of the Royal Statistical Society: Series B: Statistical
  Methodology}, pages 947--972, 2015.

\bibitem[Song and Liang(2018)]{song2018nearly}
Qifan Song and Faming Liang.
\newblock Nearly optimal {Bayesian} shrinkage for high dimensional regression.
\newblock \emph{arXiv preprint arXiv:1712.08964}, 2018.

\bibitem[Touloumis(2015)]{touloumis15}
A.~Touloumis.
\newblock Nonparametric {S}tein-type shrinkage covariance matrix estimators in
  high-dimensional settings.
\newblock \emph{Comput. Statist. Data. Anal.}, 83:\penalty0 251--261, 2015.

\bibitem[Wang(2015)]{wang15}
H.~Wang.
\newblock Scaling it up: stochastic search structure learning in graphical
  models.
\newblock \emph{Bayesian Anal.}, 10\penalty0 (2):\penalty0 351--377, 2015.

\bibitem[Watson(1959)]{watson1959note}
GN~Watson.
\newblock A note on gamma functions.
\newblock \emph{Edinburgh Mathematical Notes}, 42:\penalty0 7--9, 1959.

\bibitem[Xue et~al.(2012)Xue, Ma, and Zou]{xue2012positive}
Lingzhou Xue, Shiqian Ma, and Hui Zou.
\newblock Positive-definite ℓ1-penalized estimation of large covariance
  matrices.
\newblock \emph{Journal of the American Statistical Association}, 107\penalty0
  (500):\penalty0 1480--1491, 2012.

\end{thebibliography}

\end{document}